\documentclass[a4paper,11pt]{article}
\usepackage{jheppub} % for details on the use of the package, please
\usepackage[T1]{fontenc} % if needed

\usepackage{braket}

\def\Pom{I\!\!P}
\def\Reg{I\!\!R}

\newcommand{\bp}{\mbox{\boldmath $p$}}
\newcommand{\bk}{\mbox{\boldmath $k$}}

\begin{document}


\title{\boldmath The $\rho^0$ and Drell-S\"oding contributions \\
to central exclusive production of $\pi^+ \pi^-$ pairs\\
in proton-proton collisions at high energies}

% AUTHORS:
\author[a]{Piotr Lebiedowicz,}
\author[b]{Otto Nachtmann}
\author[a,1]{and Antoni Szczurek\note{Also at Rzesz\'ow University, PL-35-959 Rzesz\'ow, Poland.}}

% AFFILIATIONS:
\affiliation[a]{Institute of Nuclear Physics PAN, PL-31-342 Krak\'ow, Poland}
\affiliation[b]{Institut f\"ur Theoretische Physik, Universit\"at Heidelberg,\\ 
Philosophenweg 16, D-69120 Heidelberg, Germany}

% E-MAILS:
\emailAdd{Piotr.Lebiedowicz@ifj.edu.pl}
\emailAdd{O.Nachtmann@thphys.uni-heidelberg.de}
\emailAdd{Antoni.Szczurek@ifj.edu.pl}

\abstract{We present a study of the central exclusive $\pi^{+} \pi^{-}$ production
via the photoproduction mechanism in nucleon-nucleon collisions.
The photon-pomeron/reggeon and pomeron/reggeon-photon exchanges 
both for the $\rho^{0}$ resonance contribution 
and the Drell-S\"oding contribution are considered.
The amplitudes for the processes are formulated in terms of vertices
respecting the standard crossing and charge-conjugation relations of Quantum Field Theory.
The coupling parameters of tensor pomeron and reggeon exchanges
are fixed based on the H1 and ZEUS experimental data for 
the $\gamma p \to \rho^{0} p$ reaction.
We present first predictions of this mechanism
for the $pp \to pp \pi^{+} \pi^{-}$ reaction being studied 
at COMPASS, RHIC, Tevatron, and LHC.
We show the influence of the experimental cuts on the integrated cross section
and on various differential distributions for outgoing particles.
Distributions in rapidities and transverse momenta of outgoing protons and pions
as well as correlations in azimuthal angle between them are presented.
We compare the photoproduction contribution to $\pi^{+} \pi^{-}$ distributions
with double pomeron/reggeon two-pion continuum.
We discuss whether the high-energy central production of $\rho^{0}$ mesons
could be selected experimentally.}

\maketitle
\flushbottom

%-------------------------------------------------------------------
\section{Introduction}
\label{sec:intro}
%-------------------------------------------------------------------

There is a growing interest in understanding 
the mechanism of exclusive $\rho^{0}$ resonance 
and continuum $\pi^+ \pi^-$ production in the nucleon-nucleon collisions.
%both from the theoretical and experimental side.
This is closely related to ongoing experimental studies of the 
COMPASS \cite{Austregesilo:2013yxa}, 
STAR \cite{Turnau_DIS2014,Adamczyk:2014ofa}, 
CDF \cite{Albrow:2013mva,Albrow_Project_new},
ALICE \cite{Schicker:2012nn,Schicker:2014aoa}, 
ATLAS \cite{Staszewski:2011bg} and 
CMS \cite{CMS_private_com,Osterberg:2014mta} collaborations.

Some time ago two of us proposed a simple Regge-like model for 
the $\pi^+ \pi^-$ continuum based on the exchange of two 
pomerons/reggeons \cite{Lebiedowicz:2009pj}.
For further work see \cite{Lebiedowicz:2011nb}.
These model studies were extended also to $K^{+}K^{-}$ production 
\cite{Lebiedowicz:2011tp}.
Predictions for experiments at different energies 
have been presented also in Chapter~2 of \cite{Lebiedowicz:thesis}
in order to make precise comparison between calculations and experimental data.
In addition to the continuum one has to include also two-pion resonances.
Production of scalar and pseudoscalar resonances was studied
by us very recently \cite{Lebiedowicz:2013ika} in the context of 
the theoretical concept of tensor pomeron proposed 
in Ref.~\cite{Ewerz:2013kda}. 
%The production of tensor resonances such as $f_2(1270)$ is a bit 
%more complicated and has not been performed yet. 
In the present paper we focus on exclusive production of 
the $\rho^{0}$ resonance followed by the decay $\rho^0 \to \pi^{+}\pi^{-}$. 
Due to its quantum numbers this resonance state cannot be produced
by pomeron-pomeron fusion. The exchanges contributing
are photon-pomeron/reggeon and reggeon-pomeron/reggeon.
%Its contribution to the $pp \to pp \pi^{+} \pi^{-}$ reaction
%was not presented so far in the literature.

The $\gamma p \to \pi^{+} \pi^{-} p$ process has been discussed 
recently \cite{Bolz:2014mya, Sauter_LowX}
within the model for tensor-pomeron and vector-odderon \cite{Ewerz:2013kda}.
It was known for a long time that the shape of the $\rho^0$
in photoproduction is skewed. An explanation was given 
by S\"oding following a suggestion by Drell
\cite{Drell:1960zz,Drell:1961zz,Soding:1965nh}; see also \cite{Szczurek:2004xe}.
The skewing is due to the interference of continuum $\pi^+ \pi^-$ production
with the $\pi^+ \pi^-$ production through the $\rho^0$ meson.
Usually there are problems of gauge invariance when 
adding these two contributions to the $\pi^+ \pi^-$ production
and restoration of the gauge invariance is to some extent arbitrary.
The authors of Ref.~\cite{Bolz:2014mya} obtained 
a gauge-invariant version of the Drell-S\"oding mechanism
which produces the skewing of the $\rho^{0}$-meson shape.
The $\gamma p \to \pi^{+}\pi^{-} p$ reaction
would be helpful to test the model and its parameters against available data.

In the literature exist some phenomenological models 
of light vector meson photoproduction in the $pp \to p V p$ reaction,
e.g. the color dipole approach \cite{Armesto:2014sma, Santos:2014vwa},
and the pQCD $k_{t}$-factorization approach 
\cite{Schafer:2007mm,Cisek:2010jk,Cisek:2011vt,Cisek:2014ala}.
In the latter case the authors also consider absorption effects
due to strong proton-proton interactions.

\begin{figure}[tbp]
\centering
(a)\includegraphics[width=0.3\textwidth]{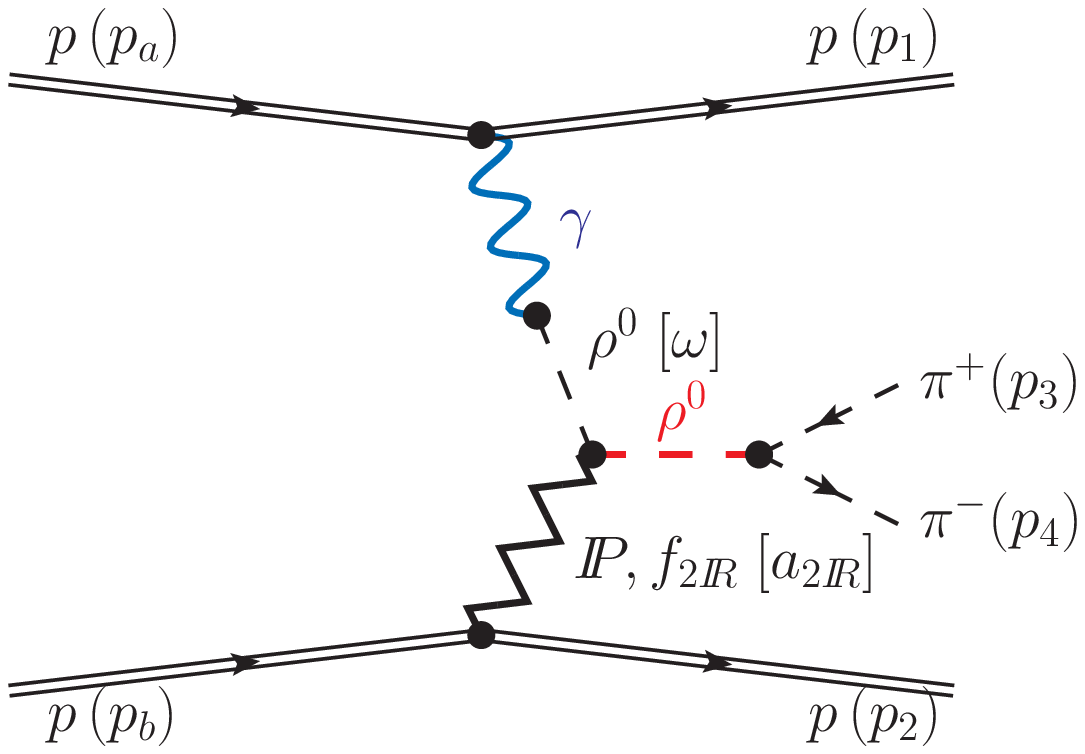}
(b)\includegraphics[width=0.3\textwidth]{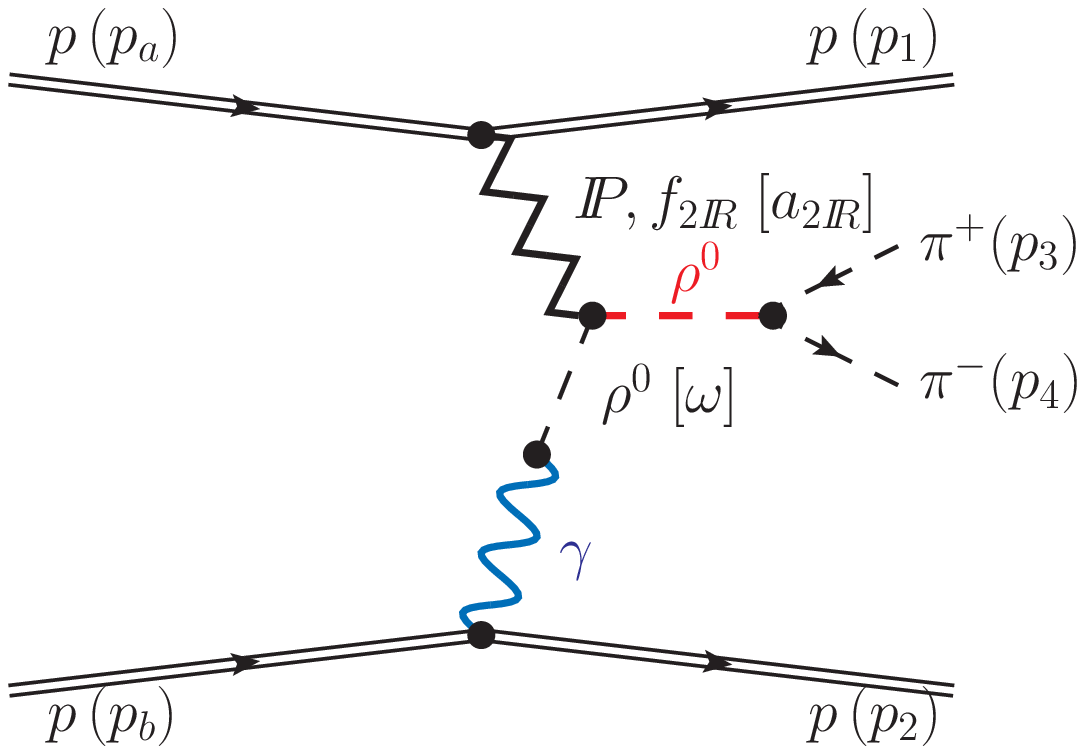}
\caption{\label{fig:gampom_pomgam_s}
The central exclusive $\rho^0$ production and its subsequent decay 
into $P$-wave $\pi^+ \pi^-$ in proton-proton collisions.}
\end{figure}

\begin{figure}[tbp]
\centering
(a)\includegraphics[width=0.28\textwidth]{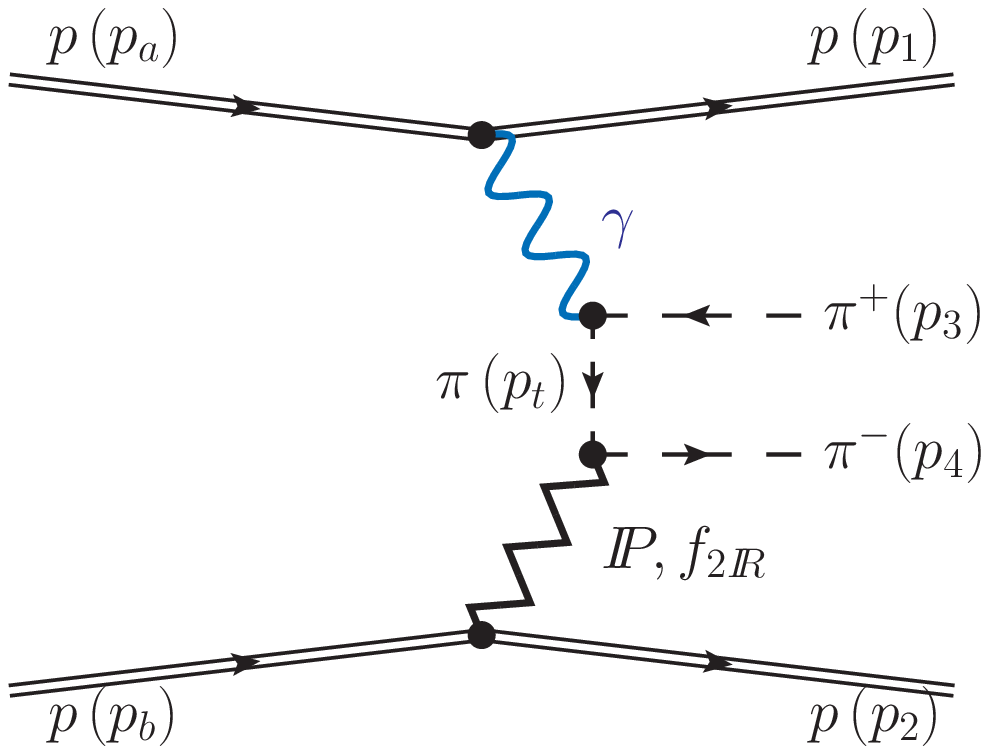}
(b)\includegraphics[width=0.28\textwidth]{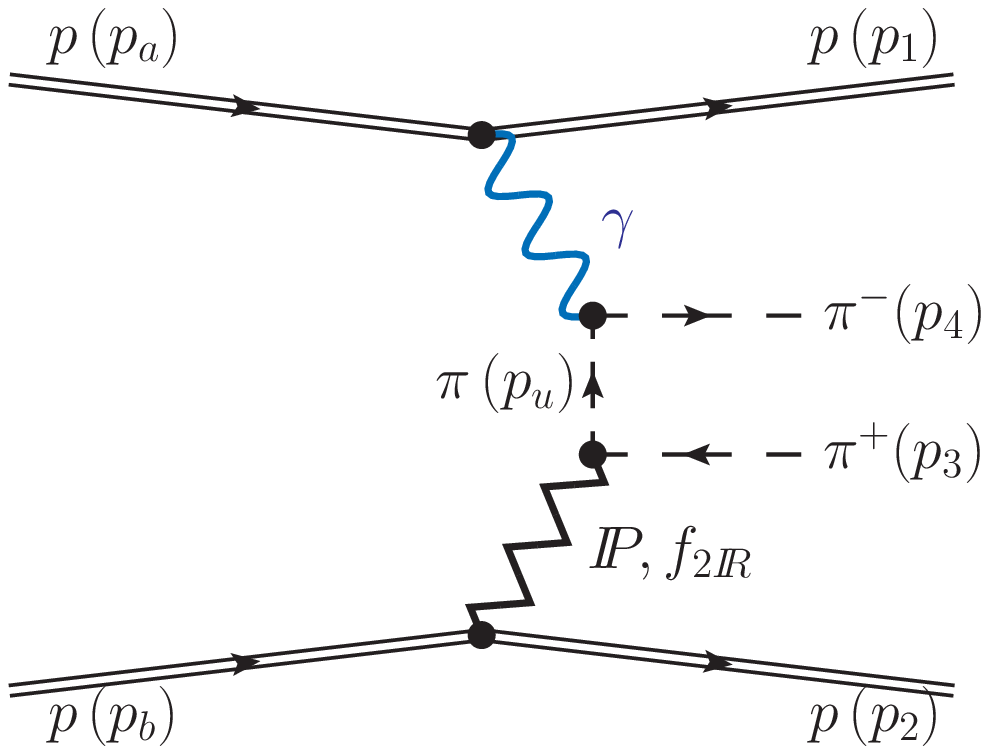}
(c)\includegraphics[width=0.28\textwidth]{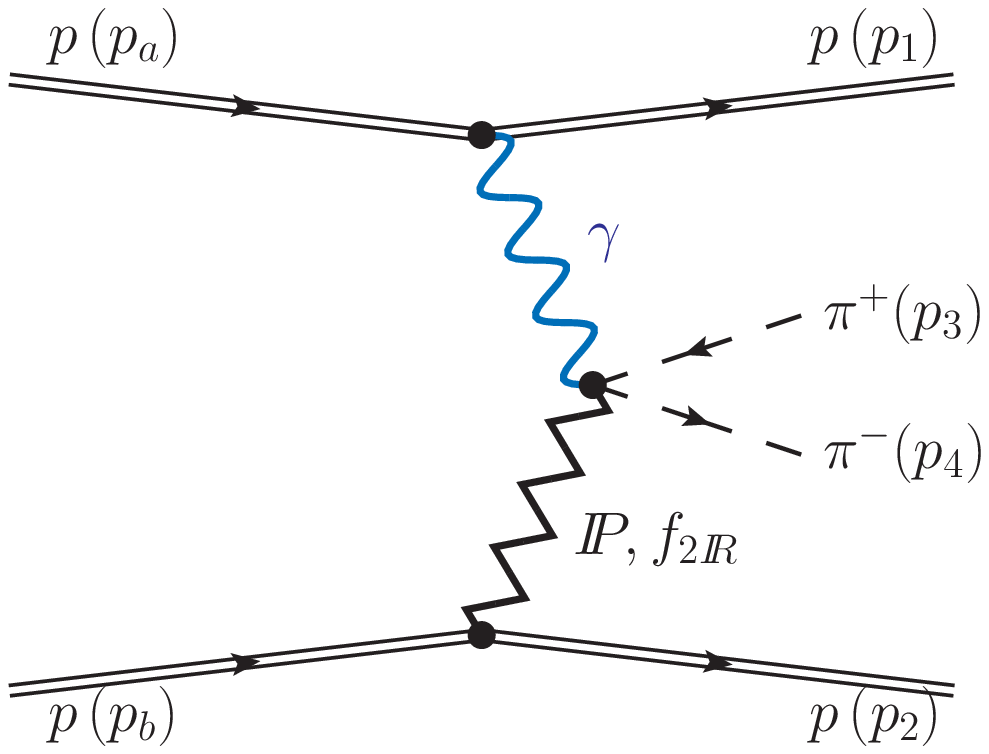}
\caption{\label{fig:gampom_pomgam_b}
The diagrams for photon-induced central exclusive continuum
$\pi^+ \pi^-$ production in proton-proton collisions.
There are also 3 additional diagrams with the role of
$(p(p_{a}), p(p_{1}))$ and $(p(p_{b}), p(p_{2}))$ exchanged.
}
\end{figure}

In this paper we focus on the four-body $pp \to pp \pi^{+}\pi^{-}$ reaction
with the pion pair produced by photon-pomeron/reggeon fusion.
For the $\rho^0$ resonance production we consider the diagrams shown in Fig.~\ref{fig:gampom_pomgam_s}.
In these diagrams all vertices and propagators will be taken here
according to Ref.~\cite{Ewerz:2013kda}.
The diagrams to be considered for the dominant non-resonant (Drell-S\"oding)
contribution are shown in Fig.~\ref{fig:gampom_pomgam_b}.
%To summarizing exclusive production of two pions in the $P$-wave 
In the following we collect formulae for the amplitudes for 
the $p p \to p p \pi^+ \pi^-$ (or $p \bar{p} \to p \bar{p} \pi^+ \pi^-$) 
reaction within the tensor pomeron model \cite{Ewerz:2013kda}.
We expect that the central exclusive $\rho^0$ photoproduction and 
its subsequent decay are the main source of $P$-wave in the $\pi^+ \pi^-$ channel 
in contrast to even waves populated in double-pomeron/reggeon processes.

A complete calculation for central exclusive $\pi^{+}\pi^{-}$
production in $pp$ collisions at high energies clearly must take 
into account more diagrams than those of Figs.~\ref{fig:gampom_pomgam_s} and \ref{fig:gampom_pomgam_b}.
For instance, in these figures we could replace
the virtual photon by the $\rho_{\Reg}$ reggeon.
As already mentioned, we can have double
pomeron/reggeon exchange leading to $\pi^{+}\pi^{-}$
in the continuum or to the $f_{2}(1270)$ resonance
decaying to $\pi^{+}\pi^{-}$.
We shall come back to all these processes in a further publication.
Here we concentrate on our first
``building block'' for this program, the processes of Figs.~\ref{fig:gampom_pomgam_s} and \ref{fig:gampom_pomgam_b}.
Due to the photon propagators occurring in these diagrams
we expect these processes to be most important
when at least one of the protons is undergoing only 
a very small momentum transfer.

There is also a non-central diffractive $\rho^{0}$ production
via the bremsstrahlung-type mechanism.
%an important background from 
%diagrams in Fig.~\ref{fig:bremsstrahlung_rho}.
But the $\rho^{0}$ mesons originating from such a mechanism
are expected to be produced very forward or very backward 
in analogy to the $\omega$-bremsstrahlung considered in Ref.~\cite{Cisek:2011vt}.
%This process is not suppressed compared to an electromagnetic coupling.
%Thus, it could still be important for $\pi^+ \pi^-$ production even in 
%the central region. 
Similar processes were discussed at high energies also for the exclusive $\pi^{0}$ meson \cite{Lebiedowicz:2013vya}, and $\gamma$ \cite{Lebiedowicz:2013xlb} production.

Our paper is organised as follows.
In section~\ref{sec:section_2} we discuss the $\gamma p \to \rho^{0} p$ reaction.
Turning to the $pp \to pp \pi^{+} \pi^{-}$ reaction we give
in section~\ref{sec:section_3} analytic expressions 
for the non-resonant (Drell-S\"oding) and the resonant (through the $\rho^{0}$ meson) amplitudes.
In section~\ref{sec:section_4} we present numerical results 
for total and differential cross sections
and discuss interference effects between the two contributions.
Moreover, we will present our prediction for the two-pion invariant mass
distribution at LHC energy of 7~TeV in proton-proton collisions, 
which is currently under analysis by the ALICE and CMS collaborations.

Closely related to the reaction $pp \to pp \pi^{+} \pi^{-}$
studied by us here are the reactions of central $\pi^{+} \pi^{-}$
production in ultra-peripheral nucleon-nucleus
and nucleus-nucleus collisions, $pA \to pA \pi^{+} \pi^{-}$
and $AA \to AA \pi^{+} \pi^{-}$.
For the latter process high-energy data exist from 
the STAR \cite{Adler:2002sc,Abelev:2007nb,Abelev:2008ew,Agakishiev:2011me}
and the ALICE \cite{Nystrand:2014vra} collaborations.
For theoretical reviews treating such collisions
see for instance \cite{Baur:2001jj,Bertulani:2005ru,Baltz:2007kq}.
The application of our methods, based on the tensor-pomeron concept,
to collisions involving nuclei is an interesting problem
which, however, goes beyond the scope of the present work.

%--------------------------------------------------------
%--------------------------------------------------------

%--------------------------------------------------------
%--------------------------------------------------------

%-------------------------------------------------------------------
\section{Photoproduction of $\rho^{0}$ meson}
\label{sec:section_2}
%-------------------------------------------------------------------

\begin{figure}[tbp]
\centering
%(a)\includegraphics[width=0.3\textwidth]{gam_s0.eps}
\includegraphics[width=0.35\textwidth]{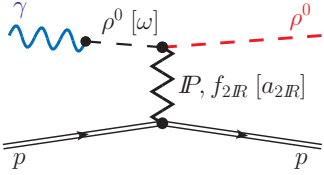}
%(c)\includegraphics[width=0.3\textwidth]{gam_s2.eps}
\caption{\label{fig:photoprod_rho_diagrams}
Photoproduction of a $\rho^0$ meson and its subsequent decay 
into a $\pi^+ \pi^-$ pair via pomeron and subleading reggeon exchanges.}
\end{figure}

The amplitude for the $\gamma p \to \rho^{0} p$ reaction
shown in Fig.~\ref{fig:photoprod_rho_diagrams}
includes not only pomeron ($\Pom$), but also reggeon 
($f_{2 \Reg}$, $a_{2 \Reg}$) exchanges.
%--------------------------------------------------------
%--------------------------------------------------------

First, we write down the amplitude for the $\gamma p \to \rho^{0} p$ reaction
via the tensor-pomeron exchange as follows
\begin{eqnarray}
&&\Braket{\rho^{0}(p_{\rho},\lambda_{\rho}),p(p_{2},\lambda_{2})
|{\cal T}|
\gamma(q,\lambda_{\gamma}),p(p_{b},\lambda_{b})} 
\equiv  
\nonumber \\  
&& {\cal M}_{\lambda_{\gamma} \lambda_{b} \to \lambda_{\rho} \lambda_{2}} =
(-i) \, (\epsilon^{(\rho)\,\mu})^* \,
i\Gamma_{\mu \nu \alpha \beta}^{(\Pom \rho \rho)}(p_{\rho},q) \,  
i\Delta^{(\rho)\,\nu \kappa}(q) \,
i\Gamma^{(\gamma \to \rho)}_{\kappa \sigma}(q)\,
\epsilon^{(\gamma)\,\sigma} 
%c^{(\gamma \to \rho)}\,
%(\Delta_{T}^{(\rho)})^{-1} \,
\nonumber \\ 
&& \qquad \qquad \qquad \quad \times i\Delta^{(\Pom)\,\alpha \beta, \delta \eta}(s,t)\, 
\bar{u}(p_{2}, \lambda_{2}) 
i\Gamma_{\delta \eta}^{(\Pom pp)}(p_{2},p_{b}) 
u(p_{b}, \lambda_{b}) \,,
\label{rhop_tot_opt_aux0}
\end{eqnarray}
where $p_{b}$, $p_{2}$ and $\lambda_{b}$, $\lambda_{2} = \pm \frac{1}{2}$ 
denote the four-momenta and helicities of the ingoing and outgoing protons,
$\epsilon^{(\gamma)}$ and $\epsilon^{(\rho)}$ are the polarisation vectors 
for photon and $\rho^{0}$ meson with the four-momenta $q$, $p_{\rho}$
and helicities $\lambda_{\gamma} = \pm 1$, $\lambda_{\rho} = \pm 1, 0$, respectively.
We use standard kinematic variables
\begin{eqnarray}
&&s = W_{\gamma p}^{2} = (p_{b} + q)^{2} = (p_{2} + p_{\rho})^{2}\,, \nonumber \\
&&t = (p_{2} - p_{b})^{2} = (p_{\rho} -q)^{2}\,.
\label{2to2_kinematic}
\end{eqnarray}

The $\Pom \rho \rho$ vertex is given in \cite{Ewerz:2013kda} by 
formula (3.47).
The propagator of the tensor-pomeron exchange is written as
(see (3.10) of \cite{Ewerz:2013kda}):
\begin{eqnarray}
i \Delta^{(\Pom)}_{\mu \nu, \kappa \lambda}(s,t) = 
\frac{1}{4s} \left( g_{\mu \kappa} g_{\nu \lambda} 
                  + g_{\mu \lambda} g_{\nu \kappa}
                  - \frac{1}{2} g_{\mu \nu} g_{\kappa \lambda} \right)
(-i s \alpha'_{\Pom})^{\alpha_{\Pom}(t)-1}
\label{pomeron_propagator}
\end{eqnarray}
and fulfils the following relations
\begin{eqnarray}
&&\Delta^{(\Pom)}_{\mu \nu, \kappa \lambda}(s,t) = 
\Delta^{(\Pom)}_{\nu \mu, \kappa \lambda}(s,t) =
\Delta^{(\Pom)}_{\mu \nu, \lambda \kappa}(s,t) =
\Delta^{(\Pom)}_{\kappa \lambda, \mu \nu}(s,t) \,,\nonumber \\
&&g^{\mu \nu} \Delta^{(\Pom)}_{\mu \nu, \kappa \lambda}(s,t) = 0, \quad 
g^{\kappa \lambda} \Delta^{(\Pom)}_{\mu \nu, \kappa \lambda}(s,t) = 0 \,.
\label{pomeron_propagator_aux}
\end{eqnarray}
Here the pomeron trajectory $\alpha_{I\!\!P}(t)$ is assumed to be 
of standard linear form with intercept slightly above 1:
\begin{eqnarray}
\alpha_{\Pom}(t) = \alpha_{\Pom}(0)+\alpha'_{\Pom}\,t, \quad 
\alpha_{\Pom}(0) = 1.0808,\quad \alpha'_{\Pom} = 0.25 \; \mathrm{GeV}^{-2}\,.
\label{pomeron_trajectory}
\end{eqnarray}
%
%soft pomeron parameters: $\alpha_{\Pom}(0) = 1.0808$
%and $\alpha'_{\Pom} = 0.25$~GeV$^{-2}$ will be used in practical calculations. 
The corresponding coupling of tensor pomeron to protons (antiprotons)
including a vertex form-factor, taken here to be
the Dirac electromagnetic form factor of the proton $F_{1}(t)$ for simplicity 
(see Section~3.2 of \cite{Donnachie:2002en}),
is written as (see (3.43) of \cite{Ewerz:2013kda}):
\begin{eqnarray}
&&i\Gamma_{\mu \nu}^{(\Pom pp)}(p',p)= 
i\Gamma_{\mu \nu}^{(\Pom \bar{p} \bar{p})}(p',p)
\nonumber\\
&& \qquad =-i 3 \beta_{\Pom NN} F_{1}\bigl((p'-p)^2\bigr)
\left\lbrace 
\frac{1}{2} 
\left[ \gamma_{\mu}(p'+p)_{\nu} 
     + \gamma_{\nu}(p'+p)_{\mu} \right]
- \frac{1}{4} g_{\mu \nu} ( p\!\!\!/' + p\!\!\!/ )
\right\rbrace , \qquad \;\;\;
%\nonumber\\
\label{vertex_pomNN}
\end{eqnarray}
where $p\!\!\!/ = \gamma_{\kappa} p^{\kappa}$ and $\beta_{\Pom NN} = 1.87$~GeV$^{-1}$.
A sufficiently good representation of the Dirac form factor $F_{1}(t)$ is given by the dipole formula
\begin{eqnarray}
F_{1}(t)= \frac{4 m_{p}^{2}-2.79\,t}{(4 m_{p}^{2}-t)(1-t/m_{D}^{2})^{2}} \,,
\label{Fpomproton}
\end{eqnarray}
where $m_{p}$ is the proton mass and $m_{D}^{2} = 0.71$~GeV$^{2}$
is the dipole mass squared.

For the $f_{2 \Reg}$ reggeon exchange a similar form of the propagator 
and the $f_{2 \Reg}pp$ effective vertex is assumed, 
see (3.12) and (3.49) of \cite{Ewerz:2013kda}, with the Regge parameters:
\begin{eqnarray}
\alpha_{\Reg_{+}}(t) = \alpha_{\Reg_{+}}(0)+\alpha'_{\Reg_{+}}\,t, \quad 
\alpha_{\Reg_{+}}(0) = 0.5475,\quad \alpha'_{\Reg_{+}} = 0.9 \; \mathrm{GeV}^{-2}\,.
\label{reggeon_trajectory}
\end{eqnarray}
%
%$\alpha_{f_2}(0)$ = 0.5475 and $\alpha'_{f_{2 \Reg}} = 0.9$~GeV$^{-2}$.
The $f_{2 \Reg}pp$ vertex is obtained from (\ref{vertex_pomNN})
replacing $3 \beta_{\Pom NN}$ by $g_{f_{2 \Reg} pp}/M_{0}$
with $M_{0} = 1$~GeV and $g_{f_{2 \Reg} pp} = 11.04$.

In the high-energy small-angle approximation we get, 
using (D.19) in Appendix~D of \cite{Lebiedowicz:2013ika},
%when we use:
%
%\begin{eqnarray}
%\bar{u}(p_{2}, \lambda_{2}) 
%i\Gamma_{\kappa \lambda}^{(\Pom pp)}(p_{2},p_{b}) 
%u(p_{b}, \lambda_{b}) \cong 
%2 (p_2+p_b)^{\kappa} (p_2+p_b)^{\lambda} \, \delta_{\lambda_{2} \lambda_{b}} \,,
%\label{rhop_tot_opt_aux2}
%\end{eqnarray}
%
%one gets:
%
\begin{eqnarray}
{\cal M}_{\lambda_{\gamma} \lambda_{b} \to \lambda_{\rho} \lambda_{2}} (s, t) \cong
%c^{(\gamma \to \rho)}\,
&&i e \dfrac{m_{\rho}^{2}}{\gamma_{\rho}}\,
\Delta_{T}^{(\rho)}(0) \,
%i\Delta^{(\rho)}_{\sigma \mu}(0)
%\epsilon_{\gamma}^{\mu} \,
%(\epsilon_{\rho}^{\nu})^* \,
(\epsilon^{(\rho)\, \mu})^*
\epsilon^{(\gamma)\, \nu} 
V_{\mu \nu \kappa \lambda}(s,t,q,p_{\rho})
\nonumber \\
&& \times
2(p_2+p_b)^{\kappa} (p_2+p_b)^{\lambda} \,
\delta_{\lambda_{2} \lambda_{b}} F_{1}(t) F_{M}(t) \,.
\label{rhop_tot_opt_aux}
\end{eqnarray}
Here 
%$c^{(\gamma \to \rho)} = i e m_{\rho}^{2}/\gamma_{\rho}$, 
$4 \pi/ \gamma_{\rho}^{2} =  0.496$,
$(\Delta_{T}^{(\rho)}(0))^{-1} = -m_{\rho}^{2}$
with the meson mass $m_{\rho}$ 
%and decay width $\Gamma_{\rho, tot}$ 
taken from PDG \cite{Agashe:2014kda}.
The function $V_{\mu \nu \kappa \lambda}(s,t,q,p_{\rho})$ has the form
\begin{eqnarray}
&&V_{\mu \nu \kappa \lambda}(s,t,q,p_{\rho})=\frac{1}{4s}
\nonumber \\
&& \times \bigg\{
2 \Gamma_{\mu \nu \kappa \lambda}^{(0)}(p_{\rho},-q)
\left[
3 \beta_{\Pom NN} \, a_{\Pom \rho\rho} 
(- i s \alpha'_{\Pom})^{\alpha_{\Pom}(t)-1}
+ M_0^{-1} g_{f_{2 \Reg}pp} \, a_{f_{2 \Reg} \rho \rho}
(- i s \alpha'_{\Reg_{+}})^{\alpha_{\Reg_{+}}(t) -1} 
\right]
\nonumber \\
&& -\Gamma_{\mu \nu \kappa \lambda}^{(2)}(p_{\rho},-q)
\left[
3 \beta_{\Pom NN} \, b_{\Pom \rho \rho} 
(- i s \alpha'_{\Pom} )^{\alpha_{\Pom}(t)-1} 
+ M_0^{-1} g_{f_{2 \Reg}pp} \, b_{f_{2 \Reg} \rho \rho}
(- i s \alpha'_{\Reg_{+}})^{\alpha_{\Reg_{+}}(t) -1} 
\right] \bigg\}\,,\nonumber \\
\label{rhop_tot_opt_aux2}
\end{eqnarray}
where the explicit tensorial functions $\Gamma_{\mu \nu \kappa \lambda}^{(i)}(p_{\rho},-q)$, 
$i$ = 0, 2,
are given in Ref.~\cite{Ewerz:2013kda}, formulae (3.18) and (3.19), respectively.
%The $f_{2 \Reg} pp$ coupling is fixed from phenomenology \cite{Ewerz:2013kda}:
%$g_{f_{2 \Reg} pp}= 11.04$ and $M_{0} = 1$~GeV.
%Two rank-four tensor functions 
%are defined in \cite{Ewerz:2013kda}, see formulas (3.18) and (3.19), respectively.
In (\ref{rhop_tot_opt_aux}) $F_{M}(t)$ is the pion electromagnetic form factor 
in a parametrization valid for $t < 0$,
\begin{eqnarray}
F_{M}(t)=\frac{1}{1-t/\Lambda_{0}^{2}}\,,
\label{F_pion}
\end{eqnarray}
where $\Lambda_{0}^{2} = 0.5$~GeV$^{2}$; see e.g.~(3.22) of \cite{Donnachie:2002en}
and (3.34) of \cite{Ewerz:2013kda}.

\begin{figure}[tbp]
\centering
\includegraphics[width=0.48\textwidth]{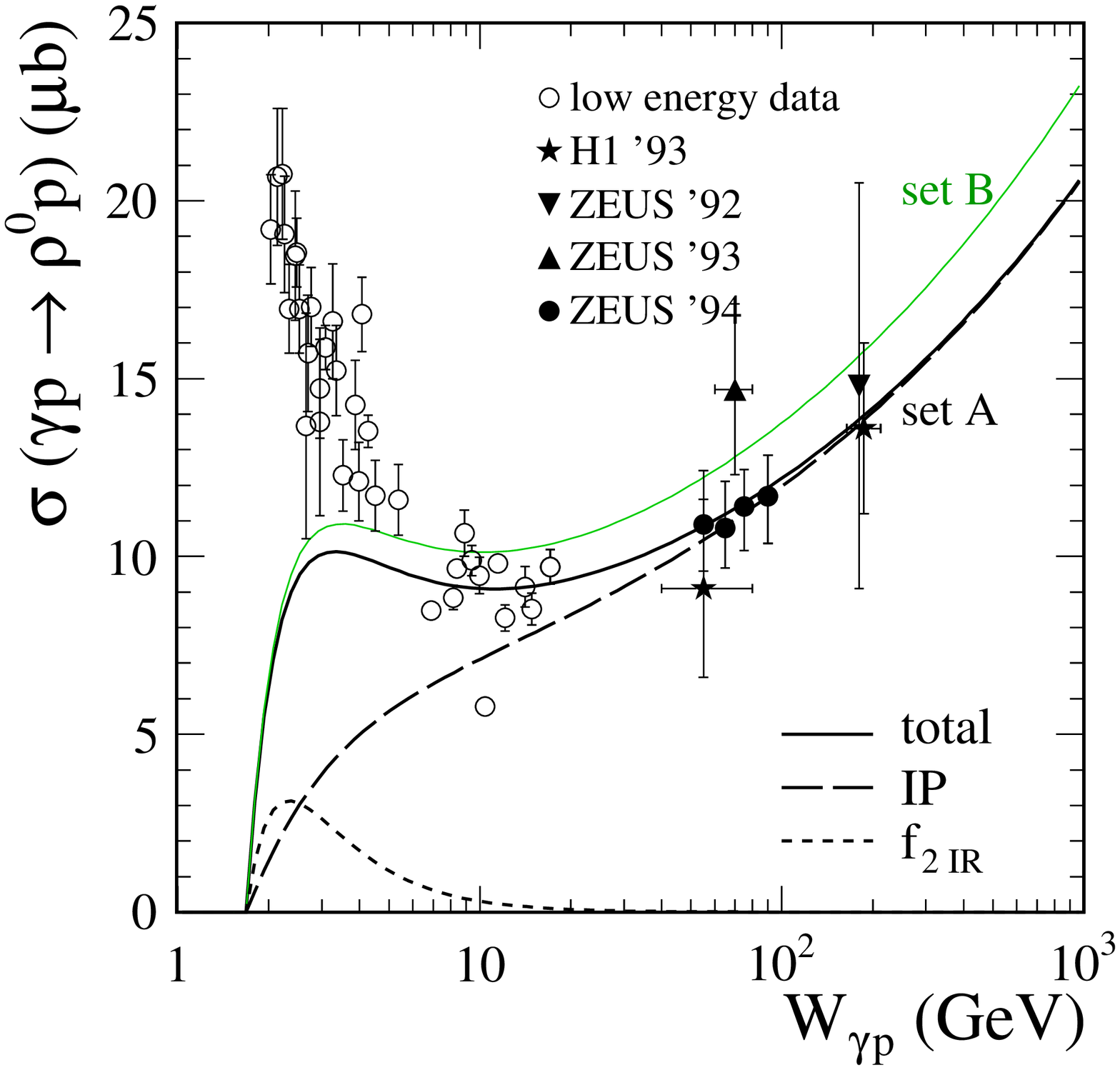}
\includegraphics[width=0.48\textwidth]{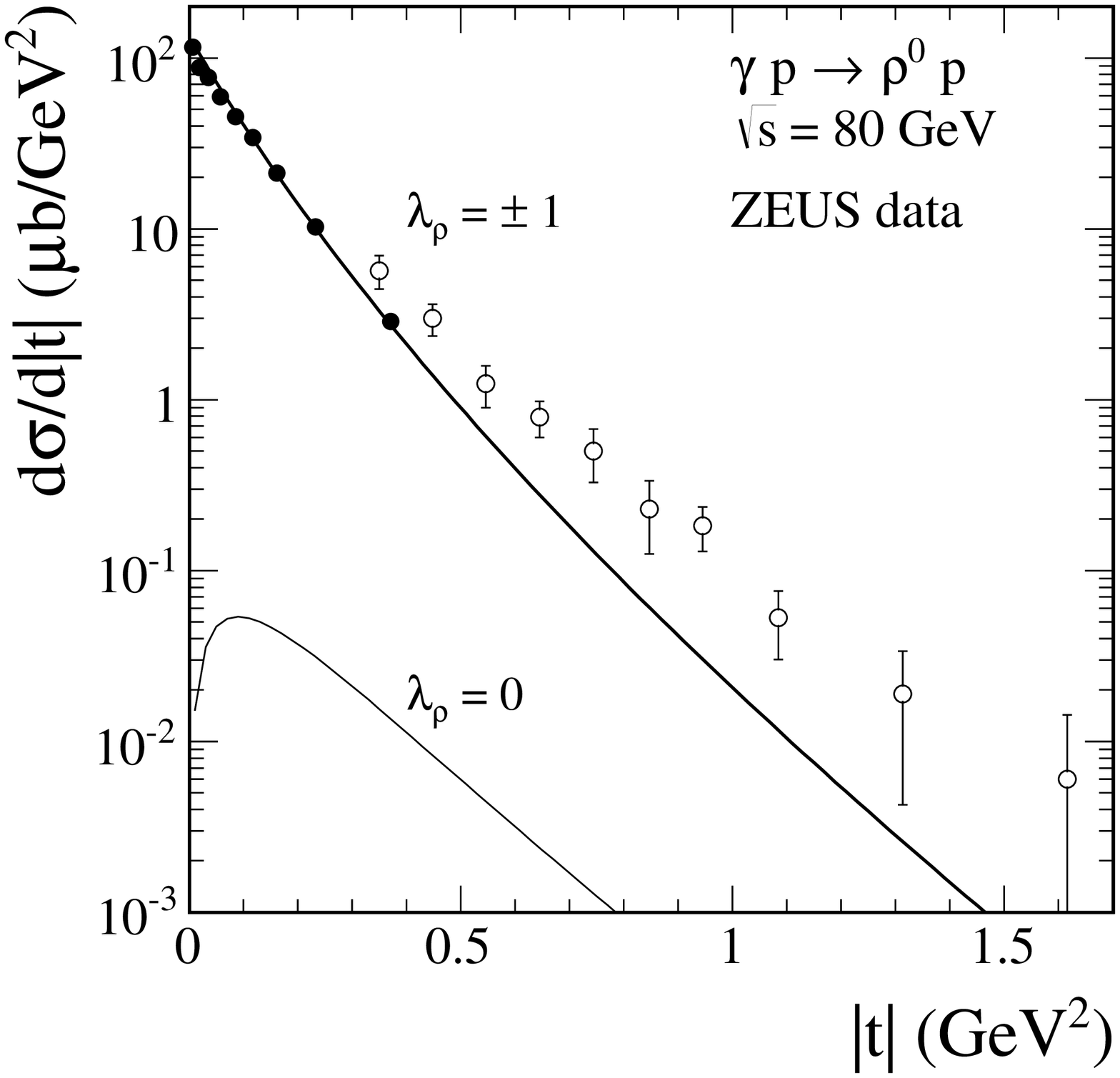}
\caption{\label{fig:photoprod_rho2}
Left panel:
The elastic $\rho^{0}$ photoproduction cross section
as a function of the center-of-mass energy $W_{\gamma p}$.
Our results are compared with the HERA data
\cite{Derrick:1994dt,Derrick:1995vq,Derrick:1996yt,Aid:1996bs,Breitweg:1997ed} (solid marks) 
and with a compilation of low energy data (open circles); 
see Fig.~10 of Ref.~\cite{Breitweg:1997ed} for more references.
The solid line corresponds to results with both the pomeron and $f_{2 \Reg}$ exchanges.
The individual pomeron and reggeon exchange contributions
denoted by the long-dashed and short-dashed lines, respectively, are presented.
The black lines represent parameter set~A of coupling constants given by (\ref{setA}) 
%both for the $\Gamma^{(0)}$ and $\Gamma^{(2)}$ tensor functions,
while the green line represents set~B (\ref{setB}).
%for the $\Gamma^{(2)}$ function alone.
%We take the $\Pom \rho \rho$ and $f_{2 I\!\!R} \rho \rho$ coupling constants
%$a_{\Pom \rho \rho} = 0.45$~GeV$^{-3}$,
%$b_{\Pom \rho \rho} = 6.50$~GeV$^{-1}$ and
%$a_{f_{2 \Reg} \rho \rho} = 2.91$~GeV$^{-3}$, 
%$b_{f_{2 \Reg} \rho \rho} = 5.80$~GeV$^{-1}$, respectively.
%The blue line represents the result with the $\Gamma^{(2)}$ tensor function only 
%and for the coupling constants given by (\ref{setII}).
Right panel: 
The differential cross section $d\sigma/d|t|$ for the 
$\gamma p \to \rho^{0} p$ process.
The ZEUS data at low $|t|$ (at $\gamma p$ average center-of-mass 
energy $<\sqrt{s}> = 71.7$~GeV \cite{Breitweg:1997ed})
and at higher $|t|$ (at $<\sqrt{s}> = 94$~GeV \cite{Breitweg:1999jy}) are shown.
The top and bottom lines represent the results at $\sqrt{s} = 80$~GeV
for $\rho^{0}$ meson transverse ($\lambda_{\rho} = \pm 1$)
and longitudinal ($\lambda_{\rho} = 0$) polarisation, respectively.
Here we have used parameter set~A of coupling constants given by (\ref{setA}).
%Bottom panel: For $\Gamma^{(2)}$ (two solid lines) and $\Gamma^{(0)}$ 
%(two dashed lines) couplings.
%For longitudinal and transverse $\lambda_{\rho}$ polarisation (top line) and 
%with transverse $\lambda_{\rho}$ only (bottom line).
}
\end{figure}

Fig.~\ref{fig:photoprod_rho2} (left panel) shows 
the integrated cross section for the $\gamma p \to \rho^{0} p$ reaction,
calculated from (\ref{rhop_tot_opt_aux}),
as function of the center-of-mass energy together with the experimental data.
We have checked that the cross section from the $a_{2 \Reg}$-exchange
is about three orders of magnitude smaller than from the $f_{2 \Reg}$-exchange.
This is due to the fact, that the $\gamma$-$\omega$ coupling is much smaller than the 
$\gamma$-$\rho$ coupling and $g_{a_{2 \Reg} pp} \ll g_{f_{2 \Reg} pp}$;
see (3.50) and (3.52) of \cite{Ewerz:2013kda} .
Thus, in the present calculations we have neglected the 
$a_{2 \Reg}$-exchange contribution.
As shown in \cite{Ewerz:2013kda} the $\Pom \rho \rho$ and $f_{2 \Reg} \rho \rho$ coupling constants $a$ and $b$ are expected to approximately fulfil the relations:
\footnote{The coupling constants of the leading trajectories in Eq.~(\ref{rhop_tot_opt_aux2})
have been estimated from the parametrization of total cross sections
for $\pi^{+} p$ and $\pi^{-} p$ scattering assuming
%
%\begin{eqnarray}
$\sigma_{tot}(\rho^{0} (\epsilon^{(\lambda_{\rho})}), p)
= \frac{1}{2}
\left[ \sigma_{tot}(\pi^{+}, p) + \sigma_{tot}(\pi^{-}, p) \right]$
%, \quad 
%\mathrm{for} \; 
for $\lambda_{\rho} = \pm 1$.
%\,,
%\label{rhop_tot_opt_aux}
%\end{eqnarray}
%
The theoretical formulae of total $\pi p$ cross sections
are discussed in Section~7.1 of \cite{Ewerz:2013kda}.}
%which fulfils the relations from:
%\footnote{but some approximations have been done here.}
%
\begin{eqnarray}
&& 2 m_{\rho}^{2} \, a_{\Pom \rho \rho} 
                   + b_{\Pom \rho \rho}
     = 4 \beta_{\Pom \pi \pi} = 7.04 \; \mathrm{GeV}^{-1} \,,
\label{rhop_tot_opt_aux_pom}\\ 
&& 2 m_{\rho}^{2} \, a_{f_{2 \Reg} \rho \rho} 
                   + b_{f_{2 \Reg} \rho \rho}
     = M_{0}^{-1} g_{f_{2 \Reg} \pi \pi} = 9.30 \; \mathrm{GeV}^{-1} \,;
\label{rhop_tot_opt_aux_reg}
\end{eqnarray}
see (7.27) and (7.28) of \cite{Ewerz:2013kda}.
In our calculations two parameter sets of coupling constants are used:
% to fit the $\gamma p \to \rho^{0} p$ data:
%
\begin{eqnarray}
\mathrm{set\;A:}  \;
&& a_{\Pom \rho \rho} = 0.7 \; \mathrm{GeV}^{-3}\,, 
a_{f_{2 \Reg} \rho \rho} = 0 \; \mathrm{GeV}^{-3}\,, \nonumber\\
&& b_{\Pom \rho \rho} = 6.2 \; \mathrm{GeV}^{-1}\,,
b_{f_{2 \Reg} \rho \rho} = 9.3 \; \mathrm{GeV}^{-1}\,,
\label{setA}\\ 
\mathrm{set\;B:} \;
&& a_{\Pom \rho \rho} = a_{f_{2 \Reg} \rho \rho} = 0 \; \mathrm{GeV}^{-3}\,,
b_{\Pom \rho \rho} = 7.04 \; \mathrm{GeV}^{-1}\,,
b_{f_{2 \Reg} \rho \rho} = 9.3 \; \mathrm{GeV}^{-1}\,.
\label{setB}
\end{eqnarray}
%
%The parameters of set~A (\ref{setA}) fulfil the relations 
%(\ref{rhop_tot_opt_aux_pom}) and (\ref{rhop_tot_opt_aux_reg}).
%Note that the result represents the parameters of set~A (\ref{setA}),
%see black solid line,
%more accurately describes the experimental data
%than the result obtained with only the $\Gamma^{(2)}$ tensor function (set~B, (\ref{setB})),
%see blue solid line.
%
%for a larger value of $b_{f_{2 \Reg} \rho \rho}$ coupling
%but breaks the above relation (\ref{rhop_tot_opt_aux_reg}).}
At low energies there are other processes contributing, such as
meson exchanges (e.g. $\pi^{0}$, $\eta$, $\sigma$), the $\rho^{0}$~bremsstrahlung,
baryonic resonances decaying into the $\rho^{0} p$ channel etc.
Thus, the Regge terms should not be expected to fit the low-energy data precisely.
We refer the reader to \cite{Friman:1995qm,Laget:2000gj,Oh:2003gm,Oh:2003aw,Riek:2008ct,Obukhovsky:2009th}
for reviews and details concerning the $\rho^{0}$ photoproduction mechanism at low energies.
We see from Fig.~\ref{fig:photoprod_rho2}, left panel, that our model
calculation describes the total cross section
for $\gamma p \to \rho^{0} p$ fairly well for energies $W_{\gamma p} \gtrsim 8$~GeV.

In the right panel of Fig.~\ref{fig:photoprod_rho2} 
we show the differential cross section 
for elastic $\rho^{0}$ photoproduction.
The calculations, performed for $\sqrt{s} = 80$~GeV,
are compared with ZEUS data \cite{Breitweg:1997ed, Breitweg:1999jy}.
We can see that, according to our calculation,
the amplitude for longitudinal $\rho^{0}$ meson polarisation
is negligible and vanishes at $t = 0$.
The sum of the $\lambda_{\rho} = \pm 1$ contributions describes the data well
up to $|t| \approx 0.5$~GeV$^{2}$.
%At very low energies an agreement with data is not expected as
%further processes, not included in our high-energy model, contribute to the cross section.
%The above relations (\ref{setA}) and (\ref{setB}) are based 
%on the assumptions that $\sigma_{tot}(\gamma p)$ behaves,
%apart from the VDM factor as $\frac{1}{2} \left( \sigma_{tot}(\pi^{+} p) + \sigma_{tot}%(\pi^{-} p) \right)$.
%Clearly, these relations need not to be fulfilled exactly.
%But they should indicate the correct order of magnitude.
%\textbf{It would be interesting to see how a fit to $\sigma_{tot}(\gamma p)$ works.}

%--------------------------------------------------------
%--------------------------------------------------------

In Ref.~\cite{Bolz:2014mya} detailed model calculations for the reaction
$\gamma p \to \pi^{+} \pi^{-} p$ have been presented
using the approach to soft scattering from \cite{Ewerz:2013kda}.
The diagrams considered in \cite{Bolz:2014mya} include $\rho^{0}$ production,
as in our Fig.~\ref{fig:photoprod_rho_diagrams}, 
but also a number of other processes;
see Fig.~1 of \cite{Bolz:2014mya}. Our results here for $\rho^{0}$ production
are in agreement with those from \cite{Bolz:2014mya};
see Figs.~3 and 4 there. This gives a very valuable cross check
of the programs used for the calculations in \cite{Bolz:2014mya}
and in our present paper.
We shall now go on and calculate with the same methods amplitudes 
and cross sections corresponding to the diagrams of 
Figs.~\ref{fig:gampom_pomgam_s} and \ref{fig:gampom_pomgam_b}.

%-----------------------------------------------------------------------
\section{The central exclusive two-pion production}
\label{sec:section_3}
%-----------------------------------------------------------------------

We shall study exclusive production of $\pi^+ \pi^-$ 
in proton-proton collisions at high energies
\begin{eqnarray}
p(p_{a},\lambda_{a}) + p(p_{b},\lambda_{b}) \to
p(p_{1},\lambda_{1}) + \pi^{+}(p_{3}) + \pi^{-}(p_{4}) + p(p_{2},\lambda_{2}) \,,
\label{2to4_reaction}
\end{eqnarray}
where $p_{a,b}$, $p_{1,2}$ and $\lambda_{a,b}$, $\lambda_{1,2} = \pm \frac{1}{2}$ 
denote the four-momenta and helicities of the protons, 
and $p_{3,4}$ denote the four-momenta of the charged pions, respectively.
In the following we will calculate the contributions from the diagrams
of Figs.~\ref{fig:gampom_pomgam_s} and \ref{fig:gampom_pomgam_b}
to the process (\ref{2to4_reaction}), that is,
the photon-pomeron and photon-$f_{2 \Reg}$ reggeon exchange contributions.
These processes are expected to be the dominant ones for highly peripheral $pp$-collisions.
Experimentally such collision could be selected by
requiring only a very small deflection angle for 
at least one of the outgoing protons.

The kinematic variables for reaction (\ref{2to4_reaction}) are
\begin{eqnarray}
&&p_{34} = p_{3} + p_{4}, \quad q_1 = p_{a} - p_{1},  \quad q_2 = p_{b} - p_{2}, \nonumber \\
&&s = (p_{a} + p_{b})^{2} = (p_{1} + p_{2} + p_{34})^{2}, \quad M_{\pi \pi}^{2} = p_{34}^{2},\nonumber \\
&&s_{1} = (p_{a} + q_{2})^{2} = (p_{1} + p_{34})^{2},  \quad 
s_{2} = (p_{b} + q_{1})^{2} = (p_{2} + p_{34})^{2}, \nonumber \\
&&t_1 = q_{1}^{2},  \quad t_2 = q_{2}^{2}\,;
\label{2to4_kinematic}
\end{eqnarray}
see also Appendix~D of \cite{Lebiedowicz:2013ika}.

The Born amplitude for exclusive photoproduction of $\pi^+ \pi^-$ 
% $\rho^0$ resonance 
%in the $P$-wave 
can be written as the following sum:
\begin{eqnarray}
{\cal M}^{Born}_{pp \to pp\pi^{+}\pi^{-}} =
{\cal M}^{(\gamma \Pom)} +
{\cal M}^{(\Pom \gamma)} +
{\cal M}^{(\gamma f_{2 \Reg})} +
{\cal M}^{(f_{2 \Reg} \gamma)} \,.
\label{2to4_reaction_pp}
\end{eqnarray}
If we want to treat $p \bar{p}$-collisions (Tevatron)
we must be careful since then there is no symmetry any more 
for the amplitude under $p(p_{a}) \leftrightarrow p(p_{b})$,
$p(p_{1}) \leftrightarrow p(p_{2})$, etc.
Using the charge conjugation ($C$) properties of the $\gamma$,
$\Pom$ and $f_{2 \Reg}$ exchanges we get for $p \bar{p}$ scattering from (\ref{2to4_reaction_pp})
%That is, in Eq.~\ref{2to4_reaction_pp} one has to change signs:
%
\begin{eqnarray}
{\cal M}^{Born}_{p \bar{p} \to p \bar{p} \pi^{+}\pi^{-}} =
{\cal M}^{(\gamma \Pom)} -
{\cal M}^{(\Pom \gamma)} +
{\cal M}^{(\gamma f_{2 \Reg})} -
{\cal M}^{(f_{2 \Reg} \gamma)} \,.
\label{2to4_reaction_ppbar}
\end{eqnarray}
Note that these sign changes are automatically obtained using
the Feynman rules for the tensor pomeron and $f_{2 \Reg}$ exchanges
but have to be implemented by hand for the vectorial pomeron and $f_{2 \Reg}$ exchanges.

%-------------------------------------------------------------------
\subsection{$\rho^{0}$-resonance contribution}
%-------------------------------------------------------------------

The ``bare'' amplitude (excluding rescattering effects) 
for the $\gamma \Pom$-exchange, see
diagram~(a) in Fig.~\ref{fig:gampom_pomgam_s},
can be written in terms of our building blocks as follows:
\begin{eqnarray}
&&{\cal M}^{(\gamma \Pom)}_{\lambda_{a} \lambda_{b} \to \lambda_{1} \lambda_{2} \pi^{+}\pi^{-}} 
= (-i)
\bar{u}(p_{1}, \lambda_{1}) 
i\Gamma^{(\gamma pp)}_{\mu}(p_{1},p_{a}) 
u(p_{a}, \lambda_{a}) \nonumber \\
&&\qquad \times  
i\Delta^{(\gamma)\,\mu \sigma}(q_{1})\, 
i\Gamma^{(\gamma \to \rho)}_{\sigma \nu}(q_{1})\,
%c^{(\gamma \to \rho)} \,
i\Delta^{(\rho)\,\nu \rho_{1}}(q_{1}) \,
i\Delta^{(\rho)\,\rho_{2} \kappa}(p_{34})\,
i\Gamma^{(\rho \pi \pi)}_{\kappa}(p_{3},p_{4})
\nonumber \\
&& \qquad \times 
i\Gamma^{(\Pom \rho \rho)}_{\rho_{2} \rho_{1} \alpha \beta}(p_{34},q_{1})\, 
i\Delta^{(\Pom)\,\alpha \beta, \delta \eta}(s_{2},t_{2}) \,
\bar{u}(p_{2}, \lambda_{2}) 
i\Gamma^{(\Pom pp)}_{\delta \eta}(p_{2},p_{b}) 
u(p_{b}, \lambda_{b}) \,.
%\sum_{\mu \nu \alpha \beta \gamma \delta \kappa \lambda}
%&&V_{1,em}^{\mu}(p_a, p_1) \left( \frac{-i g_{\mu \nu} }{t_1} \right)
%V_{(a)}^{\nu \kappa \alpha \beta}(q_1, q_2)
%\Delta^{(\Pom)}_{\alpha \beta \gamma \delta} \left(p_b, p_2 \right)
%\Gamma^{\gamma \delta}_{{2},\Pom}\left( p_b, p_2 \right) 
%\nonumber \\
%&&\Delta^{\rho^0}_{\kappa \lambda}(s_{34}) 
%V_{\rho^0 \to \pi^+ \pi^-}^{\lambda}(p_3,p_4)  \; ,
\label{amplitude_gamma_pomeron}
\end{eqnarray}
%
%\begin{eqnarray}
%{\cal M}^{\Pom \gamma} 
%= \sum_{\mu \nu \alpha \beta \gamma \delta \kappa \lambda}
%&&V_{2,em}^{\mu}(p_b, p_2) \left( \frac{-i g_{\mu \nu} }{t_2} \right)
%V_{(b)}^{\nu \kappa \alpha \beta}(q_1, q_2)
%\Delta^{(\Pom)}_{\alpha \beta \gamma \delta} \left(p_a, p_1 \right)
%\Gamma^{\gamma \delta}_{{1},\Pom}\left( p_a, p_1 \right) 
%\nonumber \\
%&&\Delta^{\rho^0}_{\kappa \lambda}(s_{34})
%V_{\rho^0 \to \pi^+ \pi^-}^{\lambda}(p_3,p_4)  \; .
%\label{amplitude_pomeron_gamma}
%\end{eqnarray}
%
All propagators and vertices used in (\ref{amplitude_gamma_pomeron})
are defined in section~3 of \cite{Ewerz:2013kda};
see also Appendix~B of \cite{Bolz:2014mya}.
For the $\Pom \gamma$-exchange the amplitude has the same structure
with $p(p_{a}), p(p_{1}) \leftrightarrow p(p_{b}), p(p_{2})$,
$t_{1} \leftrightarrow t_{2}$ and $s_{2} \leftrightarrow s_{1}$.
%where $s_{1} = (p_{a} + q_{2})^{2} = (p_{1} + p_{34})^{2}$ 
%or $s_{2} = (p_{b} + q_{1})^{2} = (p_{2} + p_{34})^{2}$,
%$p_{34} = p_{3} + p_{4}$
%and $t_1 = (p_{1} - p_{a})^{2}$ or $t_2 = (p_{2} - p_{b})^{2}$ in the proper subsystem.
In a similar way we can write down the $\gamma f_{2 \Reg}$ 
and $f_{2 \Reg} \gamma$ amplitudes.

For simplicity, in the following we shall 
consider the amplitude (\ref{amplitude_gamma_pomeron}) 
in the high-energy small-angle limit; see Appendix~D of 
\cite{Lebiedowicz:2013ika}.
Including both, pomeron and $f_{2 \Reg}$ exchanges, we obtain in this way
\begin{eqnarray}
&&{\cal M}^{(\gamma \Pom+\, \gamma f_{2 \Reg})}_{\lambda_{a} \lambda_{b} \to \lambda_{1} \lambda_{2} \pi^{+}\pi^{-}} 
\simeq i e (p_1 + p_a)^{\mu} F_1(t_1) \delta_{\lambda_{1}\lambda_{a}}
%\bar{u}(p_{1}, \lambda_{1}) 
%i\Gamma_{\mu}^{(\gamma pp)}(p_{1},p_{a}) 
%u(p_{a}, \lambda_{a}) 
\nonumber \\
&&\qquad \times  
%\Delta^{(\gamma)\,\mu \nu}(q_{1})\, 
%c^{(\gamma \to \rho)} \,
e \frac{m_{\rho}^{2}}{\gamma_{\rho}} \frac{1}{t_{1}}\,
\Delta^{(\rho)}_{\mu \rho_{1}}(q_{1}) \,
\Delta^{(\rho)}_{\rho_{2} \kappa}(p_{34})\,
\frac{g_{\rho \pi \pi}}{2} (p_{3} - p_{4})^{\kappa} \,
%F_{\rho \pi \pi}(s_{34}) F_{\rho \pi \pi}(s_{34})
\tilde{F}^{(\rho)}(q_{1}^{2}) \tilde{F}^{(\rho)}(p_{34}^{2})
\nonumber \\
&& \qquad \times 
V^{\rho_{2} \rho_{1} \alpha \beta}(s_{2}, t_{2}, q_{1}, p_{34}) F_{M}(t_{2}) \,
2 (p_2 + p_b)_{\alpha} (p_2 + p_b)_{\beta} F_1(t_2) 
\delta_{\lambda_{2}\lambda_{b}}
%i\Delta^{(\Pom)\,\alpha \beta, \delta \eta}(s_{2},t_{2}) \,
%\bar{u}(p_{2}, \lambda_{2}) 
%i\Gamma_{\delta \eta}^{(\Pom pp)}(p_{2},p_{b}) 
%u(p_{b}, \lambda_{b}) 
\,.
\label{amplitude_gamma_pomeron_2}
\end{eqnarray}
%
%where the $\pm e$ sign is for $p$ and $\bar{p}$, respectively.
%The relevant structure in Eq.~(\ref{amplitude_gamma_pomeron})
%is the $\rho^{0}$ propagator times 
%the $\Gamma^{(\Pom \rho \rho)}_{\rho_{2} \rho_{1} \alpha \beta}(p_{34},q_{1})$ vertex.
The function $V^{\rho_{2} \rho_{1} \alpha \beta}(s_{2}, t_{2}, q_{1}, p_{34})$ is as defined in (\ref{rhop_tot_opt_aux2}) and
includes two tensorial functions 
$\Gamma^{(i) \, \rho_{2} \rho_{1}  \alpha \beta}(p_{34},-q_{1})$ for $i = 0,2$.
%see Eq.~(\ref{rhop_tot_opt_aux2}).
%For simplicity the electromagnetic coupling of photons to nucleons 
%can be approximated as
From (3.21) and (3.22) of \cite{Ewerz:2013kda} we have
\begin{eqnarray}
&&q_{1}^{\rho_{1}} \, \Gamma^{(i)}_{\rho_{2} \rho_{1} \alpha \beta}(p_{34},-q_{1}) = 0\,, \quad
p_{34}^{\rho_{2}} \, \Gamma^{(i)}_{\rho_{2} \rho_{1} \alpha \beta}(p_{34},-q_{1}) = 0\,, \nonumber \\
&&g^{\alpha \beta} \Gamma^{(i)}_{\rho_{2} \rho_{1} \alpha \beta}(p_{34},-q_{1}) = 0 \,, \quad i =0,2\,. 
\end{eqnarray}
Thus, the terms proportional to $q_{1}^{\mu} q_{1}^{\rho_{1}}$ in 
$\Delta^{(\rho)\,\mu \rho_{1}}(q_{1})$
cannot contribute.
The same is true for the terms proportional to $p_{34}^{\rho_{2}} p_{34}^{\kappa}$
in $\Delta^{(\rho)\,\rho_{2} \kappa}(p_{34})$.
%The $\rho^0$ meson propagator in the $t$-channel is
%taken in a simplified form:
%
%\begin{eqnarray}
%\Delta^{(\rho)\,\mu \nu}(q) = \frac{i\left(-g^{\mu \nu} [+ q^{\mu} q^{\nu} / t] \right)}
%{t - m_{\rho}^2 + i m_{\rho} \Gamma_{\rho, tot}} \,.
%\label{rho_propagator_t}
%\end{eqnarray}
%
%In Fig.~\ref{fig:rho_pipi_y_test} we show results with first term 
%(black line) and with both terms (red dashed line) in this propagator.
%The $\rho^0$ meson propagator in the $s$-channel is
%required a special comments.
%In the calculations is taken a simplified form:
%
%\begin{eqnarray}
%\Delta^{(\rho)\,\mu \nu}(p_{34}) = \frac{-ig^{\mu \nu}}
%{s_{34} - m_{\rho}^2 + i m_{\rho} \Gamma_{\rho, tot}} \,. 
%\label{rho_propagator_s}
%\end{eqnarray}
%
%Here for simplicity we completely ignored the second term in the propagator.
Thus, only the transverse part
of the $\rho^{0}$ propagator $\Delta^{(\rho)}_{T}(k^{2})$,
as defined in (4.1)-(4.4) of \cite{Ewerz:2013kda},
contributes in (\ref{amplitude_gamma_pomeron_2}).
As is easily seen, the same holds for (\ref{amplitude_gamma_pomeron}).
%The assumption going into (4.1) to (4.4) is that
%for calculating the self-energy part of the $\rho$ propagator
%it is sufficient to consider $\pi^{+}\pi^{-}$, $K^{+} K^{-}$ and $K^{0} \bar{K}^{0}$ intermediate states
%with constant coupling parameters to the $\rho$ meson.
%Physical arguments require that $\Delta^{(\rho)}_{T}(k^{2})$ must be real for $k^{2} < 4 m_{\pi}^{2}$,
%i.e. below the first kinematical threshold.
%Thus, the simple Breit-Wigner form is incorrect for $k^{2} = 0$
%since it gives an imaginary part to $\Delta^{(\rho)}_{T}(0)$.
The decay vertex for $\rho^0 \to \pi^+ \pi^-$ is well known 
(e.g. see (3.35) of \cite{Ewerz:2013kda}) 
and the relevant coupling constant is $g_{\rho \pi \pi} = 11.51$.
We emphasize that in the work \cite{Melikhov:2003hs}
the authors found a very good description of the $\rho$ line shape 
from the $e^{+} e^{-} \to \pi^{+} \pi^{-}$ reaction
- up to $k^{2} \cong 2$~GeV$^{2}$ - without a form factor in the $\rho \pi \pi$ vertex.

%\textbf{It should be advantageous to use the $\rho^{0}$ propagator
%as given in section~4.1 of \cite{Ewerz:2013kda} 
%instead of the simple Breit-Wigner expression.
%With the latter one risks to run into problems with partial-wave unitarity}.

%In our case for the $\Pom \rho \rho$ vertex the $\rho^{0}$ meson
%can be off-shell with $k^{2} \neq m_{\rho}^{2}$
In the diagram of Fig.~\ref{fig:gampom_pomgam_s}
at the pomeron-$\rho$-$\rho$ vertex the incoming $\rho$ is always off shell,
the outgoing $\rho$ also may be away from the nominal 
``mass shell'' $p_{34}^{2} = m_{\rho}^{2}$.
As suggested in \cite{Bolz:2014mya}, see (B.82) there,
we insert, therefore, in the $\Pom \rho \rho$ vertex extra form factors.
A convenient form, given in (B.85) of \cite{Bolz:2014mya} is
\begin{eqnarray}
\tilde{F}^{(\rho)}(k^{2}) 
= \left[1 + \dfrac{k^{2}(k^{2}-m_{\rho}^{2})}{\Lambda_{\rho}^{4}}  \right]^{-n_{\rho}} \,.
%\Lambda_{\rho} = 2 \,\mathrm{to}\, 5 \; \mathrm{GeV}, \; n_{\rho} = 0.4 \,\mathrm{to}\, 0.5 \,.
\label{ff_Nachtmann}
\end{eqnarray}
The form factor (\ref{ff_Nachtmann}) has the property:
$\tilde{F}^{(\rho)}(0) = \tilde{F}^{(\rho)}(m_{\rho}^{2}) = 1$
which is consistent with the traditional,
phenomenologically successful, vector-meson-dominance model.
%In our calculations presented in the Section~\ref{Results}
%we take $\Lambda_{\rho} = 2$~GeV and $n_{\rho} = 0.4$ \cite{Nachtmann_Sauter}.

%--------------------------------------------------------
%\begin{figure}[!ht]
%\includegraphics[width=0.5\textwidth]{dsig_dy_wrong.eps}
%  \caption{\label{fig:rho_pipi_y_test}
%  \small
%Rapidity distribution of pions produced from the $\rho^{0}$ decay
%at $\sqrt{s} = 7$~TeV.
%The black solid line corresponds to the $\rho^0$ meson propagator
%in $t$-channel (\ref{rho_propagator_t}) with first term in the
%propagator only, while the red dashed line represents the result with 
%both terms.
%In this calculation we have used the parameter set~A of coupling 
%constants given by (\ref{setA}).
%{\bf We do not know if the deviation of the two results at $|y_{\pi}| > 4$ is 
%of numerical origin or has a deeper origin. Your comment is welcome.}
%}
%\end{figure}
%--------------------------------------------------------

%-------------------------------------------------------------------
\subsection{Drell-S\"oding contribution}
%-------------------------------------------------------------------

The $\gamma\Pom$-exchange amplitudes
given by the diagrams shown in Fig.~\ref{fig:gampom_pomgam_b}, 
can be written on the Born level for the tensor-pomeron exchange as follows
\begin{eqnarray}
&&{\cal M}^{(a)}_{\lambda_{a} \lambda_{b} \to \lambda_{1} \lambda_{2} \pi^{+}\pi^{-}} 
= (-i)
\bar{u}(p_{1}, \lambda_{1}) 
i\Gamma^{(\gamma pp)}_{\mu}(p_{1},p_{a}) 
u(p_{a}, \lambda_{a})\,
i\Delta^{(\gamma)\,\mu \nu}(q_{1})\,
i\Gamma^{(\gamma \pi \pi)}_{\nu}(p_{t},-p_{3})\nonumber \\
&&\qquad \times  
i\Delta^{(\pi)}(p_{t}) \,
i\Gamma^{(\Pom \pi \pi)}_{\alpha \beta}(p_{4},p_{t})\,
%i\Gamma^{(\Pom \rho \rho)}_{\rho_{1} \rho_{2} \alpha \beta}(-q_{1},-p_{34})\, 
i\Delta^{(\Pom)\,\alpha \beta, \delta \eta}(s_{2},t_{2}) \,
\bar{u}(p_{2}, \lambda_{2}) 
i\Gamma^{(\Pom pp)}_{\delta \eta}(p_{2},p_{b}) 
u(p_{b}, \lambda_{b}) \,, \nonumber\\
\label{amplitude_gamma_pomeron_a}\\
%\end{eqnarray}
%
%\begin{eqnarray}
&&{\cal M}^{(b)}_{\lambda_{a} \lambda_{b} \to \lambda_{1} \lambda_{2} \pi^{+}\pi^{-}} 
= (-i)
\bar{u}(p_{1}, \lambda_{1}) 
i\Gamma^{(\gamma pp)}_{\mu}(p_{1},p_{a}) 
u(p_{a}, \lambda_{a})\,
i\Delta^{(\gamma)\,\mu \nu}(q_{1})\,
i\Gamma^{(\gamma \pi \pi)}_{\nu}(p_{4},p_{u}) \nonumber \\
&&\qquad \times  
i\Delta^{(\pi)}(p_{u}) \, 
i\Gamma^{(\Pom \pi \pi)}_{\alpha \beta}(p_{u},-p_{3})\,
%i\Gamma^{(\Pom \rho \rho)}_{\rho_{1} \rho_{2} \alpha \beta}(-q_{1},-p_{34})\, 
i\Delta^{(\Pom)\,\alpha \beta, \delta \eta}(s_{2},t_{2}) \,
\bar{u}(p_{2}, \lambda_{2}) 
i\Gamma^{(\Pom pp)}_{\delta \eta}(p_{2},p_{b}) 
u(p_{b}, \lambda_{b}) \,,  \nonumber\\
\label{amplitude_gamma_pomeron_b} \\
%\end{eqnarray}
%
%\begin{eqnarray}
&&{\cal M}^{(c)}_{\lambda_{a} \lambda_{b} \to \lambda_{1} \lambda_{2} \pi^{+}\pi^{-}} 
= (-i)
\bar{u}(p_{1}, \lambda_{1}) 
i\Gamma^{(\gamma pp)}_{\mu}(p_{1},p_{a}) 
u(p_{a}, \lambda_{a}) \,
i\Delta^{(\gamma)\,\mu \nu}(q_{1}) \nonumber \\
&&\qquad \times  
i\Gamma^{(\Pom \gamma \pi \pi)}_{\nu, \alpha \beta}(q_{1},p_{4},-p_{3}) \,
i\Delta^{(\Pom)\,\alpha \beta, \delta \eta}(s_{2},t_{2}) \,
\bar{u}(p_{2}, \lambda_{2}) 
i\Gamma^{(\Pom pp)}_{\delta \eta}(p_{2},p_{b}) 
u(p_{b}, \lambda_{b}) \,.
\label{amplitude_gamma_pomeron_c}
\end{eqnarray}
Above we have introduced $p_{t} = p_{a} - p_{1} - p_{3}$ and 
$p_{u} = p_{4} - p_{a} + p_{1}$.
In order to assure gauge invariance and ``proper'' cancellations 
among the three terms (\ref{amplitude_gamma_pomeron_a}) 
to (\ref{amplitude_gamma_pomeron_c}) 
we have introduced, somewhat arbitrarily, one common energy dependence $s_{2}$
%$\tilde{s}_{2}=(s_{24}+s_{23})/2$
for the pomeron propagator in all three diagrams 
instead of naively $s_{24}$, $s_{23}$, and $s_{2}$,
respectively, where $s_{ij} = (p_{i} + p_{j})^{2}$.
See also the discussion of this point in section~2.5 of \cite{Bolz:2014mya}
where a justification for this procedure is given.

For the $\Pom \gamma$-exchange we have the same structure as for the above amplitudes
with $p(p_{a}), p(p_{1}) \leftrightarrow p(p_{b}), p(p_{2})$,
$t_{2} \leftrightarrow t_{1}$, $s_{2} \leftrightarrow s_{1}$.
%$s_{2} \to s_{1}$, $s_{24} \to s_{13}$, 
%$s_{23} \to s_{14}$, where $s_{ij} = (p_{i} + p_{j})^{2}$.
In a similar way we can write the $\gamma f_{2 \Reg}$ and $f_{2 \Reg} \gamma$ amplitudes.

Starting with the $\Pom \pi \pi$ coupling, see Eq.~(7.3) of 
\cite{Ewerz:2013kda},
and making a minimal substitution there gives the couplings involving pions, 
photons and the pomeron (see (B.66) to (B.71) of \cite{Bolz:2014mya}).
We have the following vertex, see~Eq.(3.45) of \cite{Ewerz:2013kda}
and (B.69) of \cite{Bolz:2014mya},
\begin{eqnarray}
i\Gamma_{\alpha \beta}^{(\Pom \pi \pi)}(k',k)=
-i 2 \beta_{\Pom \pi \pi} 
\left[ (k'+k)_{\alpha}(k'+k)_{\beta} - \frac{1}{4} g_{\alpha \beta} (k' + k)^{2} \right] \, F_{M}((k'-k)^2)\,,
\label{vertex_pompipi}
\end{eqnarray}
where $\beta_{I\!\!P \pi \pi} = 1.76$~GeV$^{-1}$.
For the standard electromagnetic vertex we have
\begin{eqnarray}
i\Gamma_{\nu}^{(\gamma \pi \pi)}(k',k)=
i e (k'+k)_{\nu} \, F_{M}((k'-k)^2)\,.
\label{vertex_gampipi}
\end{eqnarray}
Finally, there is the contact term, see (B.71) of \cite{Bolz:2014mya},
\begin{eqnarray}
i\Gamma_{\nu, \alpha \beta}^{(\Pom \gamma \pi \pi)}(q,k',k)=&&
- i e 2 \beta_{\Pom \pi \pi} 
\left[  2 g_{\alpha \nu} (k'+k)_{\beta}
      + 2 g_{\beta \nu} (k'+k)_{\alpha} 
      - g_{\alpha \beta} (k'+k)_{\nu} \right] \nonumber\\
&&\times F_{M}(q^{2})\, F_{M}((k' - q - k)^{2})\,.
\label{vertex_gampompipi}
\end{eqnarray}
Compared to (B.67) and (B.71) of \cite{Bolz:2014mya}
we have introduced in both, (\ref{vertex_gampipi}) and (\ref{vertex_gampompipi}),
an extra form factor $F_{M}(q^{2})$ (\ref{F_pion}).
For an on shell photon with $q^{2} =0$ this form factor is equal to 1 and,
thus, would not make any difference 
for the calculations of photoproduction in \cite{Bolz:2014mya}.
With the normal pion propagator $i\Delta^{(\pi)}(k) = i/(k^{2}-m_{\pi}^{2})$
the above ansatz for the vertices guarantees
gauge invariance of the $\pi^{+}\pi^{-}$ continuum contribution.

In the high-energy approximation we can write for tensor-pomeron exchange
\begin{eqnarray}
&&{\cal M}^{(a)}_{\lambda_{a} \lambda_{b} \to \lambda_{1} \lambda_{2} \pi^{+}\pi^{-}} 
\simeq i e^{2} (p_1 + p_a)^{\mu}\, \delta_{\lambda_{1}\lambda_{a}}\, F_1(t_1) F_{M}(t_{1}) 
\frac{1}{t_{1}} (p_t - p_3)_{\mu} 
\frac{1}{p_{t}^{2}-m_{\pi}^{2}}\nonumber \\
&&\qquad \times  
2 \beta_{\Pom \pi \pi} \,
%\left[ 
(p_{4}+p_{t})^{\alpha}(p_{4}+p_{t})^{\beta}  
%\textcolor{red}{- \frac{1}{4} g^{\alpha \beta} (p_{4}+p_{t})^{2}} 
%\right]
\,
\frac{1}{4 s_{2}} (- i s_{2} \alpha'_{\Pom})^{\alpha_{\Pom}(t_{2})-1}
\nonumber \\
&&\qquad \times  
3 \beta_{\Pom NN}  \, 2 (p_2 + p_b)_{\alpha} (p_2 + p_b)_{\beta}\, \delta_{\lambda_{2}\lambda_{b}}\, F_1(t_2) F_{M}(t_{2}) \,, \quad
\label{amplitude_gamma_pomeron_a_approx}\\
&&{\cal M}^{(b)}_{\lambda_{a} \lambda_{b} \to \lambda_{1} \lambda_{2} \pi^{+}\pi^{-}} 
\simeq i e^{2} (p_{1} + p_{a})^{\mu} \, \delta_{\lambda_{1}\lambda_{a}}\, F_1(t_{1}) F_{M}(t_{1})
\frac{1}{t_{1}} (p_{4} + p_{u})_{\mu}
\frac{1}{p_{u}^{2}-m_{\pi}^{2}}\nonumber \\
&&\qquad \times  
2 \beta_{\Pom \pi \pi} \,
%\left[ 
(p_{u}-p_{3})^{\alpha}(p_{u}-p_{3})^{\beta} 
%\textcolor{red}{- \frac{1}{4} g^{\alpha \beta} (p_{u}-p_{3})^{2}} 
%\right] 
\,
\frac{1}{4 s_{2}} (- i s_{2} \alpha'_{\Pom})^{\alpha_{\Pom}(t_{2})-1} \nonumber \\
&&\qquad \times  
3 \beta_{\Pom NN} \,2 (p_2 + p_b)_{\alpha} (p_2 + p_b)_{\beta}\, \delta_{\lambda_{2}\lambda_{b}} \,F_1(t_2) F_{M}(t_{2}) \,, \quad
\label{amplitude_gamma_pomeron_b_approx}\\
&&{\cal M}^{(c)}_{\lambda_{a} \lambda_{b} \to \lambda_{1} \lambda_{2} \pi^{+}\pi^{-}} 
\simeq -i e^{2} (p_{1} + p_{a})^{\nu} \, \delta_{\lambda_{1}\lambda_{a}} \,
\frac{1}{t_{1}} F_1(t_{1}) F_{M}(t_{1}) \nonumber \\
&&\qquad \times  
2 \beta_{\Pom \pi \pi}\,
\left[  2 g_{\alpha \nu} (p_{4}-p_{3})_{\beta}
      + 2 g_{\beta \nu} (p_{4}-p_{3})_{\alpha}  \right]\,
\frac{1}{4 s_{2}} (- i s_{2} \alpha'_{\Pom})^{\alpha_{\Pom}(t_{2})-1}
\nonumber \\
%       \textcolor{red}{- g_{\alpha \beta} (p_{4}-p_{3})_{\nu}} \right]\nonumber \\
&&\qquad \times  
3 \beta_{\Pom NN} \,2 (p_2 + p_b)^{\alpha} (p_2 + p_b)^{\beta} \, \delta_{\lambda_{2}\lambda_{b}} \,F_1(t_2) F_{M}(t_{2}) \,.
\label{amplitude_gamma_pomeron_c_approx}
\end{eqnarray}
Gauge invariance of our expressions can be checked explicitly.
As it should be we find from (\ref{amplitude_gamma_pomeron_a_approx})
to (\ref{amplitude_gamma_pomeron_c_approx})
\begin{eqnarray}
\lbrace {\cal M}^{(a)} + {\cal M}^{(b)} + {\cal M}^{(c)} \rbrace|_{p_{1} + p_{a} \to q_{1}} = 0 \,.
\label{gauge_invariance}
\end{eqnarray}
For the $f_{2 \Reg}$-reggeon exchange the formulae have the same tensorial structure
and are obtained from (\ref{amplitude_gamma_pomeron_a_approx})
to (\ref{amplitude_gamma_pomeron_c_approx}) with the replacements
discussed in the paragraph including (\ref{reggeon_trajectory}).
The formulae above do not include hadronic form factors 
for the inner subprocesses $\gamma \Pom \to \pi \pi$.

A possible way to include form factors
for the inner subprocesses is to multiply the amplitude obtained from
(\ref{amplitude_gamma_pomeron_a_approx}) to (\ref{amplitude_gamma_pomeron_c_approx})
with a common factor, see \cite{Poppe:1986dq, Szczurek:2002bn, Klusek-Gawenda:2013rtu},
\begin{eqnarray}
{\cal M}^{(\gamma \Pom)} = ({\cal M}^{(a)} + {\cal M}^{(b)} + {\cal M}^{(c)}) \,F(p_{t}^{2},p_{u}^{2},p_{34}^{2})\,.
\label{amplitude_gamma_pomeron_abc}
\end{eqnarray}
A common form factor for all three diagrams
is chosen in order to maintain gauge invariance,
and a convenient form is
\begin{eqnarray}
F(p_{t}^{2},p_{u}^{2},p_{34}^{2}) = 
\frac{F^{2}(p_{t}^{2}) + F^{2}(p_{u}^{2})}{1 + F^{2}(-p_{34}^{2})} \,.
\label{ff_Poppe}
\end{eqnarray}
Here we take the monopole form factor 
which is normalized to unity at
the on-shell point $F(m_{\pi}^{2}) = 1$:
\begin{eqnarray}
F(p^{2}) = \frac{\Lambda_{\pi}^{2} - m_{\pi}^{2}}{\Lambda_{\pi}^{2} - p^{2}} \,.
%\exp \left( \dfrac{-(\hat{t}/\hat{u} - m_{\pi}^2)}{\Lambda_{\pi}^2} \right) \,, \quad
%F(\hat{s}) = \exp \left( \dfrac{-(\hat{s} - 4 m_{\pi}^2)}{\Lambda_{\pi}^2} \right) \,,
\label{monopole_ff}
%&&\mathrm{set\;II}: \;F(\hat{t}/\hat{u}) = 
%\exp \left( \dfrac{-(\hat{t}/\hat{u} - m_{\pi}^2)^2}{\Lambda_{\pi}^4}\right) \,, \quad
%F(\hat{s}) = \exp \left( \dfrac{-(\hat{s} - 4 m_{\pi}^2)^2}{\Lambda_{\pi}^4}\right) \,.
%\label{ff_bgrd_setB}
%&&\mathrm{set\;III}: \;F(\hat{t}/\hat{u}) = 
%\left(\hat{t}/\hat{u} - m_{\pi}^2)^2}{\Lambda_{\pi}^4}\right) \,, \quad
%F(\hat{s}) = \exp \left( \dfrac{-(\hat{s} - 4 m_{\pi}^2)^2}{\Lambda_{\pi}^4}\right) \,.
%\label{ff_bgrd_setC}
\end{eqnarray}
%
%where $p^{2}$ denote the four-momentum squared.
with $\Lambda_{\pi}$ being a free parameter.
We expect it in the range of 0.8 to 1~GeV.

%-------------------------------------------------------------------
\subsection{Absorptive corrections}
%-------------------------------------------------------------------

We should add the absorptive corrections to the Born amplitude (\ref{2to4_reaction_pp}) 
to give the full physical amplitude for the $pp \to pp \pi^{+} \pi^{-}$ reaction, 
i.e. we have
\begin{eqnarray}
{\cal {M}}_{pp \to pp \pi^{+} \pi^{-}} =
{\cal {M}}_{pp \to pp \pi^{+} \pi^{-}}^{Born} + 
{\cal {M}}_{pp \to pp \pi^{+} \pi^{-}}^{pp-rescattering}\,.
\label{amp_full}
\end{eqnarray}
Here (and above) we have for simplicity omitted 
the dependence of the amplitude on kinematic variables.
The details how to conveniently reduce the number of kinematic integration variables
are discussed in \cite{Lebiedowicz:2009pj}.

%where the auxiliary quantity $\bp_{m \perp}=\bp_{3 \perp}-\bp_{4 \perp}$.
%The $2 \to 4$ amplitude described above (\ref{amp_full})
%is used to calculate the corresponding cross sections.
%\footnote{There is also the $\pi p$ rescattering corrections
%as discussed in Ref.~\cite{Ryskin:1997zz}.}
%see e.g. Section~2.6 of \cite{Lebiedowicz:thesis}.
%The absorptive correction is often embodied in the soft gap survival probability.
The amplitude including $pp$-rescattering corrections 
in the four-body reaction discussed here is given by
\footnote{We refer the reader to 
\cite{Schafer:2007mm, Cisek:2010jk, Cisek:2011vt, Cisek:2014ala, Lebiedowicz:2013vya}
for reviews of three-body processes and details concerning the absorption corrections in the eikonal approximation
which takes into account the contribution of elastic $pp$-rescattering.
}
\begin{eqnarray}
{\cal M}_{pp \to pp \pi^{+} \pi^{-}}^{pp-rescattering}(s,\bp_{1\perp},\bp_{2\perp})=
\frac{i}{8 \pi^{2} s} \int d^{2}\bk_{\perp} {\cal M}_{pp \to pp}(s,-\bk_{\perp}^{2})
{\cal M}_{pp\to pp \pi^{+} \pi^{-}}^{Born}
(s,\tilde{\bp}_{1\perp},\tilde{\bp}_{2\perp})\,, \nonumber \\
\label{abs_correction}
\end{eqnarray}
where $\tilde{\bp}_{1\perp} = \bp_{1\perp} - \bk_{\perp}$ and
$\tilde{\bp}_{2\perp} = \bp_{2\perp} + \bk_{\perp}$.
Here, in the overall c.m. system, $\bp_{1\perp}$ and $\bp_{2\perp}$
are the transverse components of the momenta of the final-state protons
and $\bk_{\perp}$ is the transverse momentum carried around the pomeron loop.
${\cal M}_{pp \to pp}(s,-\bk_{\perp}^{2})$ 
is the elastic $pp$ scattering amplitude given by Eq.~(6.28) in \cite{Ewerz:2013kda}
for large $s$ and with the momentum transfer $t=-\bk_{\perp}^{2}$.

%-------------------------------------------------------------------
\section{Results}
\label{sec:section_4}
%-------------------------------------------------------------------

\begin{figure}[tbp]
\centering
\includegraphics[width=0.5\textwidth]{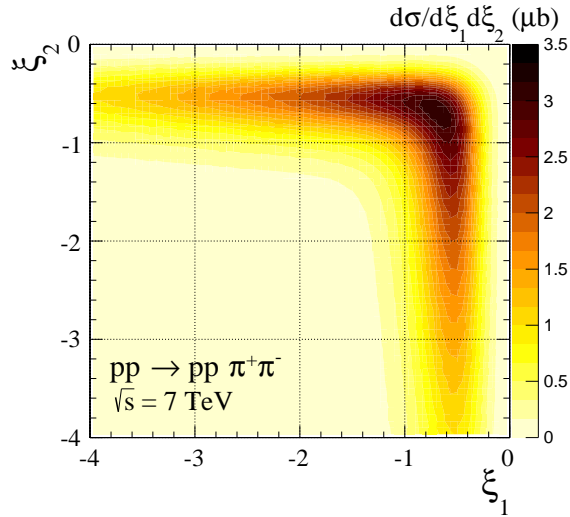}
\caption{\label{fig:map_xi1xi2}
%Differential cross section in $(y_3,y_4)$ space
Two-dimensional distribution in the auxiliary variables 
$\xi_1 = \log_{10}(p_{1\perp}/1 \,\mathrm{GeV})$ and 
$\xi_2 = \log_{10}(p_{2\perp}/1 \,\mathrm{GeV})$
for the photoproduction mechanism at $\sqrt{s} = 7$~TeV.
Plotted is $d\sigma/d\xi_{1}d\xi_{2}$ in $\mu$b.
In the calculation we have used the parameter set~A 
of coupling constants given by (\ref{setA})
and we have taken $\Lambda_{\rho} = 2$~GeV and $n_{\rho} = 0.5$ in (\ref{ff_Nachtmann}).
%at $\sqrt{s} = 200$~GeV (left panel) and $7$~TeV (right panel).
%In this calculation we have used the parameter set~B of coupling 
%constants given by (\ref{setB}).
}
\end{figure}

\begin{figure}[tbp]
\centering
\includegraphics[width=0.48\textwidth]{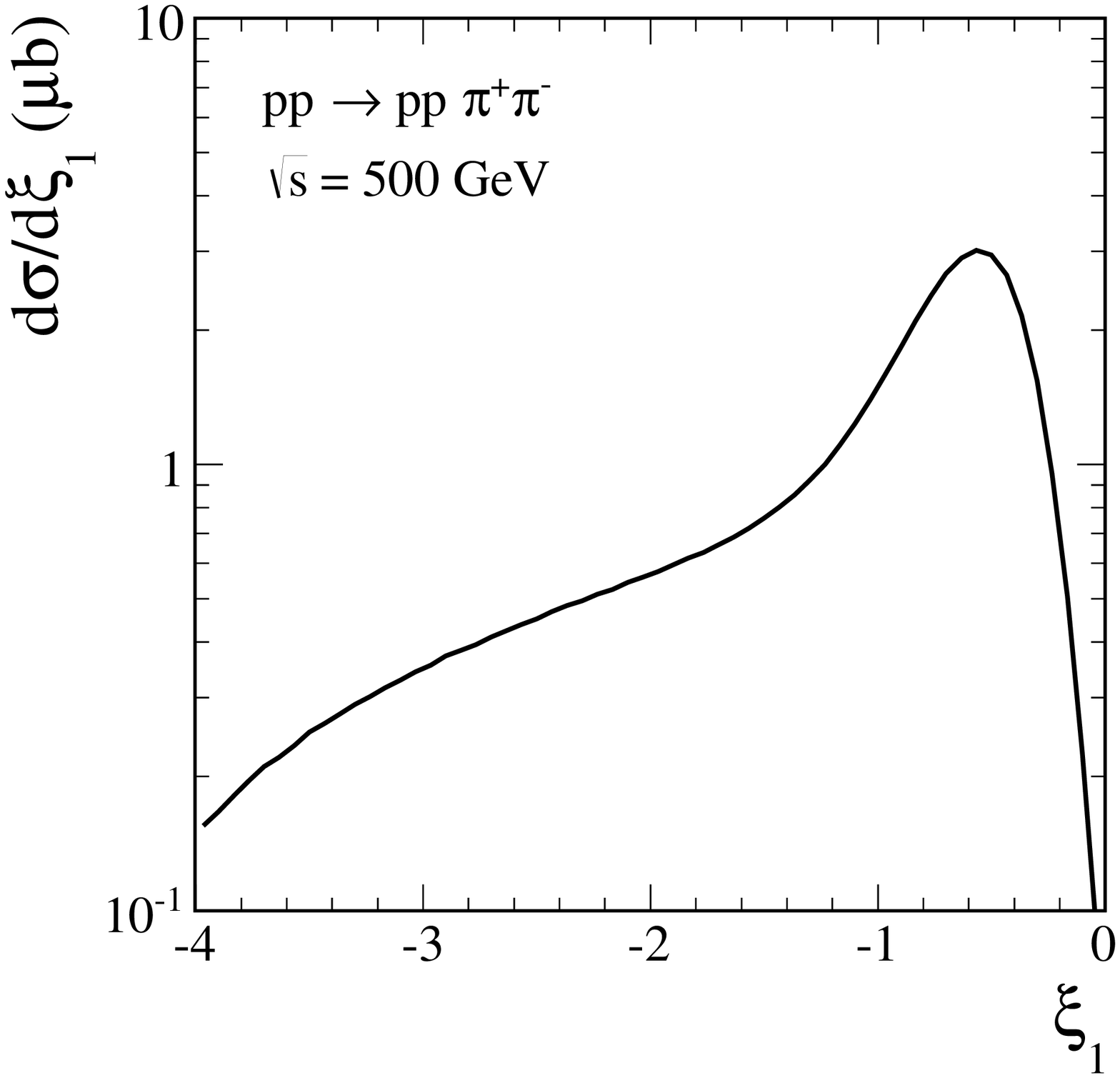}
\includegraphics[width=0.48\textwidth]{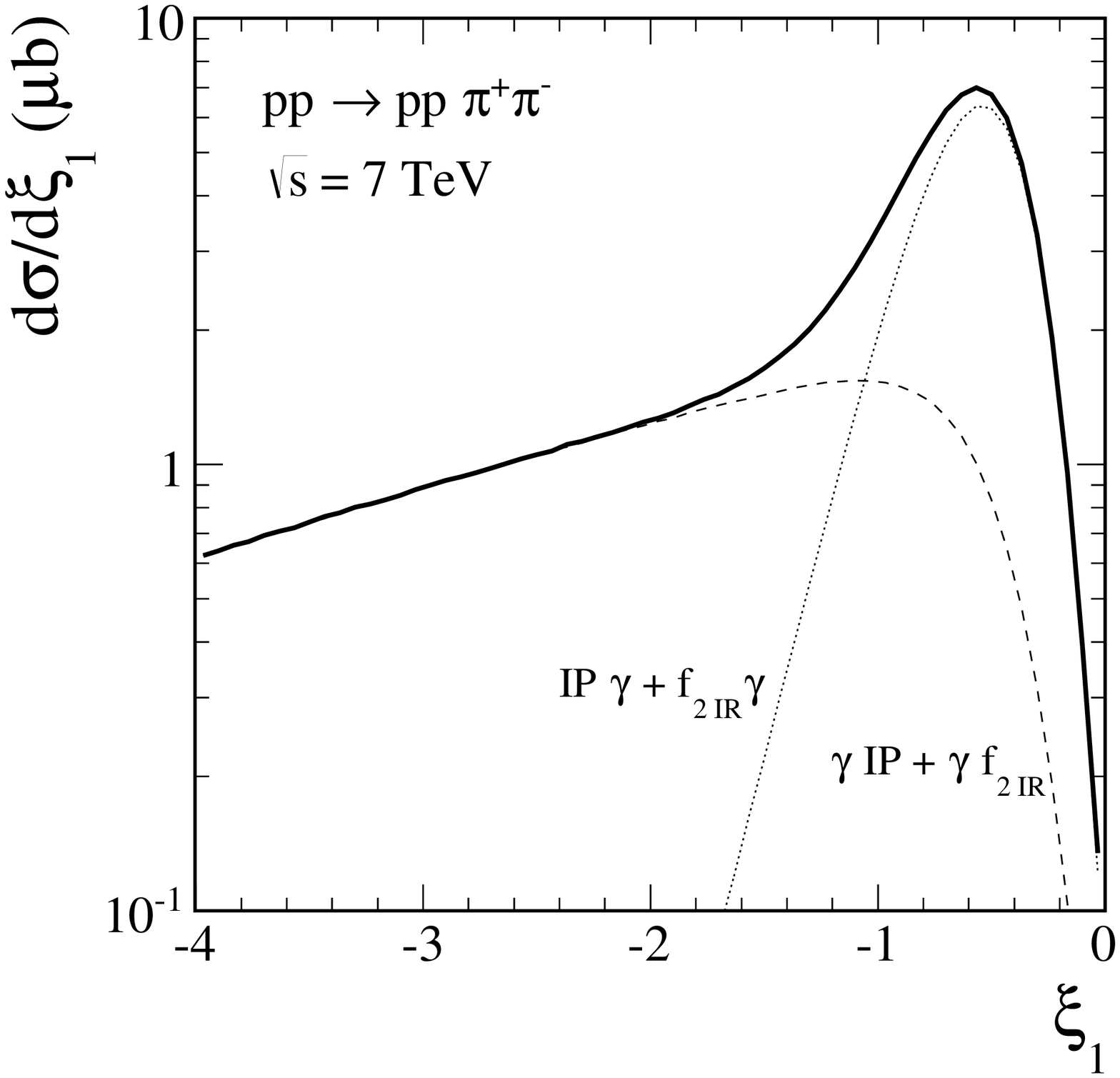}
\caption{\label{fig:rho_pipi_xi1}
The distributions in $\xi_1$ at two incident energies $\sqrt{s} = 0.5$ and $7$~TeV.
The solid line corresponds to the situation when all,
that is, the pomeron and $f_{2 \Reg}$ exchanges in the amplitude are included.
%while the long-dashed (blue) line corresponds to the pomeron exchange alone.
%The blue line is obtained for parameter set~B of coupling constants 
%given by (\ref{setB}),
%while the black lines for parameter set~A (\ref{setA}).
The short-dashed and dotted lines represent the contributions
from $\gamma-\Pom/\Reg$ and $\Pom/\Reg-\gamma$ exchanges, respectively.
}
\end{figure}

In this section we present some selected results for cross sections
for the discussed photoproduction processes.
In calculating the cross section of the four-body process (\ref{2to4_reaction})
we perform integrations in auxiliary variables
$\xi_1 = \log_{10}(p_{1\perp}/1\,\mathrm{GeV})$
and $\xi_2 = \log_{10}(p_{2\perp}/1\,\mathrm{GeV})$
instead of the outgoing proton's transverse momenta 
($p_{1\perp}$ and $p_{2\perp}$).
For example $\xi_{1} = -1$ means $p_{1\perp} = 0.1$~GeV.
The correlation between the variables $\xi_{1}$ and $\xi_{2}$ is
displayed in Fig.~\ref{fig:map_xi1xi2} at center-of-mass energy at $\sqrt{s} = 7$~TeV.
Here we include both the resonant and non-resonant contributions.
The projection on one of the axes is shown in Fig.~\ref{fig:rho_pipi_xi1} 
%we show the distributions in $\xi_1$ 
at two incident energies at $\sqrt{s} = 0.5$ and $7$~TeV.
The parameters of the model used here are indicated in the legend of
Fig.~\ref{fig:map_xi1xi2}.
%We show how the absorptive effects due to the $pp$-interaction 
%affect our distributions.
%The distributions in $\xi_{1}$ or $\xi_{2}$ are different because we have limited
%to the case of $\eta_{\gamma} > 0$ only.

%--------------------------------------------------------
%--------------------------------------------------------
%--------------------------------------------------------
%--------------------------------------------------------

\begin{figure}[tbp]
\centering
\includegraphics[width=0.48\textwidth]{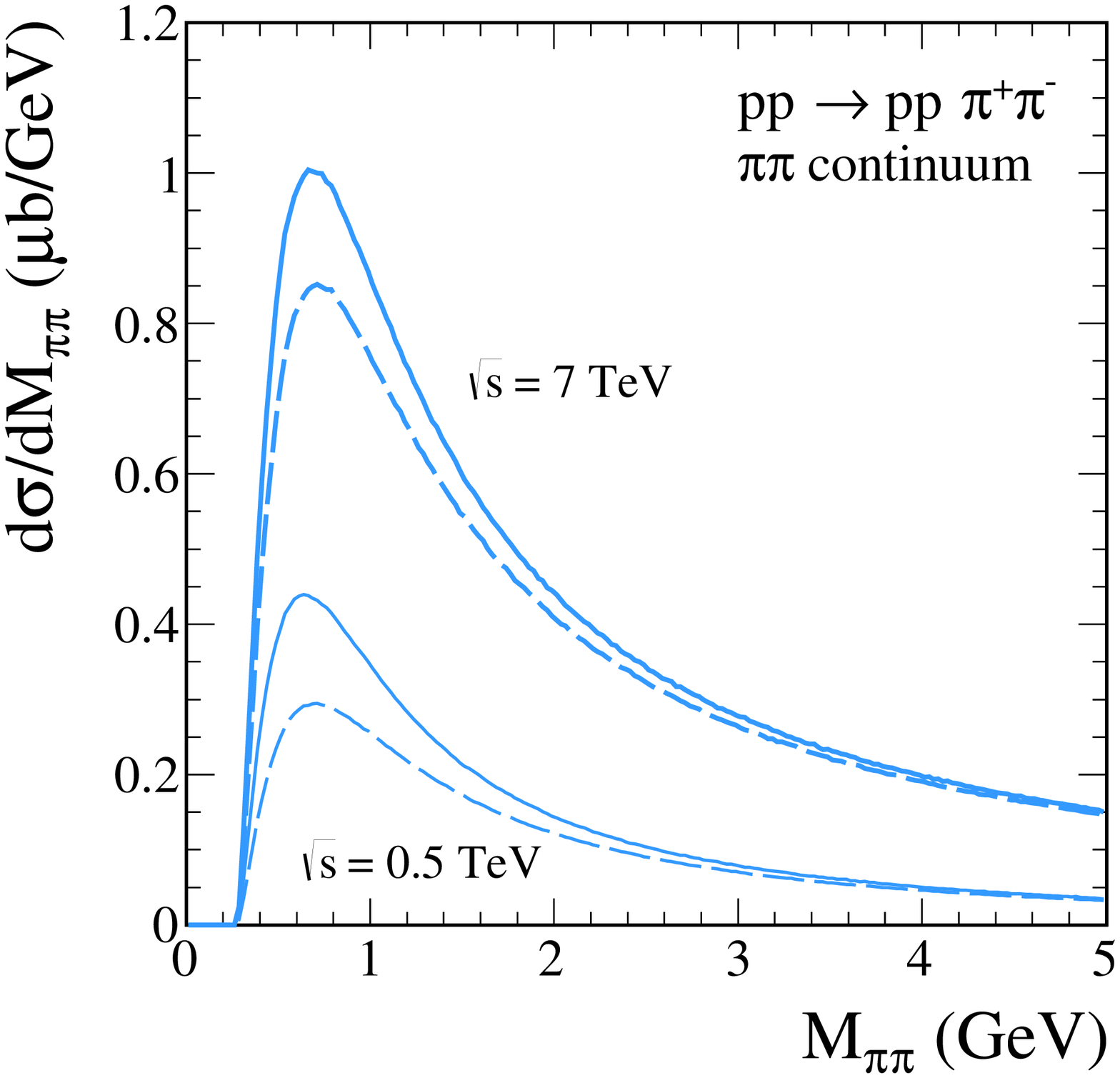}
\caption{\label{fig:rho_pipi_M34_background}
Two-pion continuum (Drell-S\"oding) contributions 
%via the tensor pomeron/$f_{2 \Reg}$ reggeon exchanges
%(blue lines) and via the vector-pomeron/reggeon exchange (violet lines) 
at $\sqrt{s}=0.5$~TeV (lower lines) and 7~TeV (upper lines).
The solid lines correspond 
to the tensor pomeron and $f_{2 \Reg}$ exchanges 
in the amplitudes while the dashed lines correspond 
to the pomeron exchange alone.
No extra form factors (\ref{ff_Poppe}) 
for inner subprocesses are included.
}
\end{figure}

In Fig.~\ref{fig:rho_pipi_M34_background} 
we show the two-pion invariant mass distributions 
for the Drell-S\"oding contribution alone 
%via the tensor pomeron/$f_{2 \Reg}$ reggeon exchanges
%the model with tensor pomeron/$f_{2 \Reg}$ reggeon exchanges.
%We wish to emphasize the different shape of the distributions for 
%the two models (the same form factor for the pomeron coupling to 
%the proton has been used).
at two c.m. energies $\sqrt{s}=0.5$ and 7~TeV,
see the lower and upper lines, respectively.
The solid lines correspond to the tensor pomeron and $f_{2 \Reg}$ 
exchanges in the amplitude while the dashed lines correspond 
to the pomeron exchange alone.
At the lower energy one can observe a larger interference effect 
between the $\gamma \Pom$ ($\Pom \gamma$) and 
the $\gamma f_{2 \Reg}$ ($f_{2 \Reg} \gamma$) components 
in the amplitude (\ref{2to4_reaction_pp}).
%--------------------------------------------------------
%--------------------------------------------------------

\begin{figure}[tbp]
\centering
\includegraphics[width=0.48\textwidth]{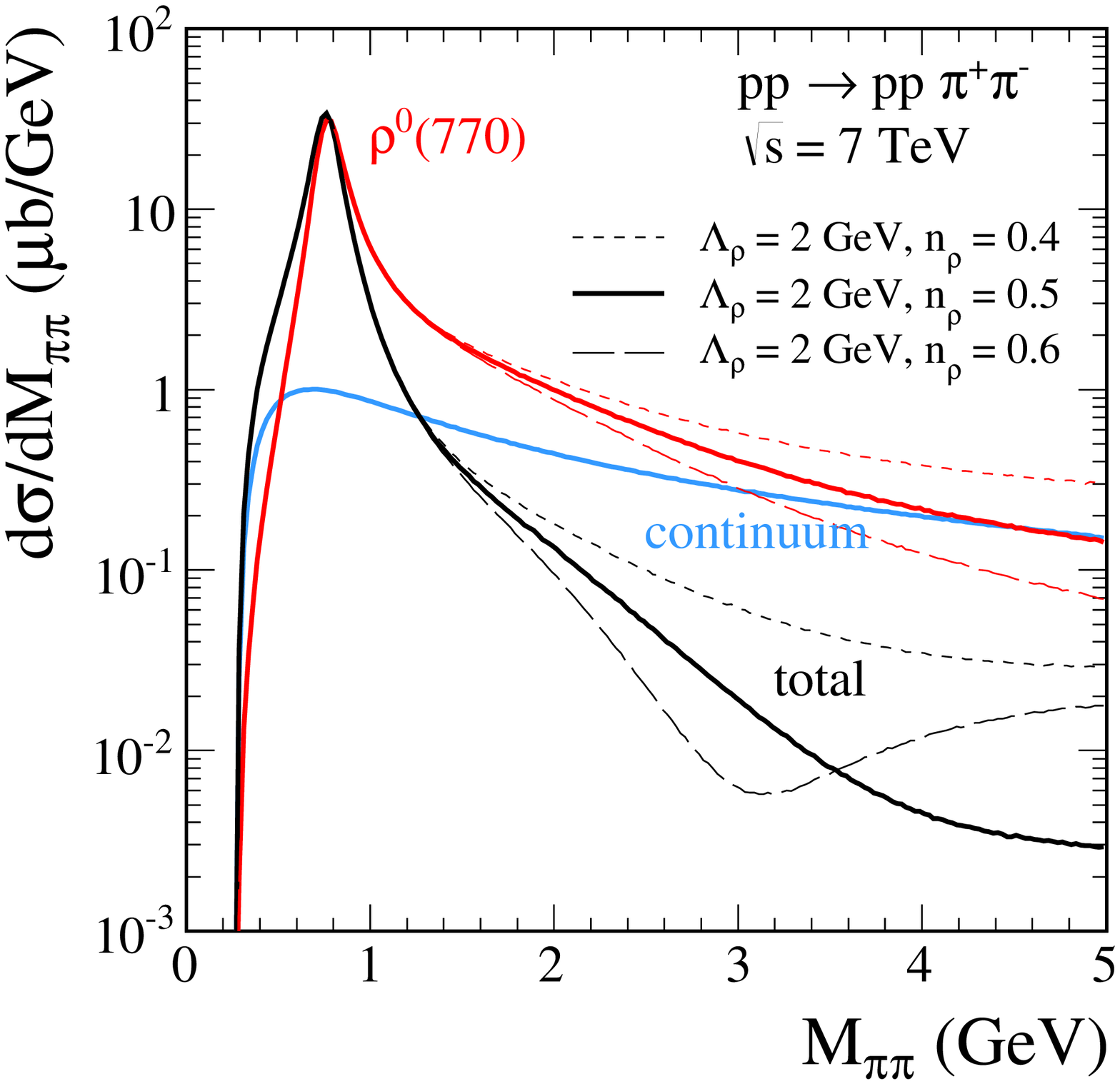}
\includegraphics[width=0.48\textwidth]{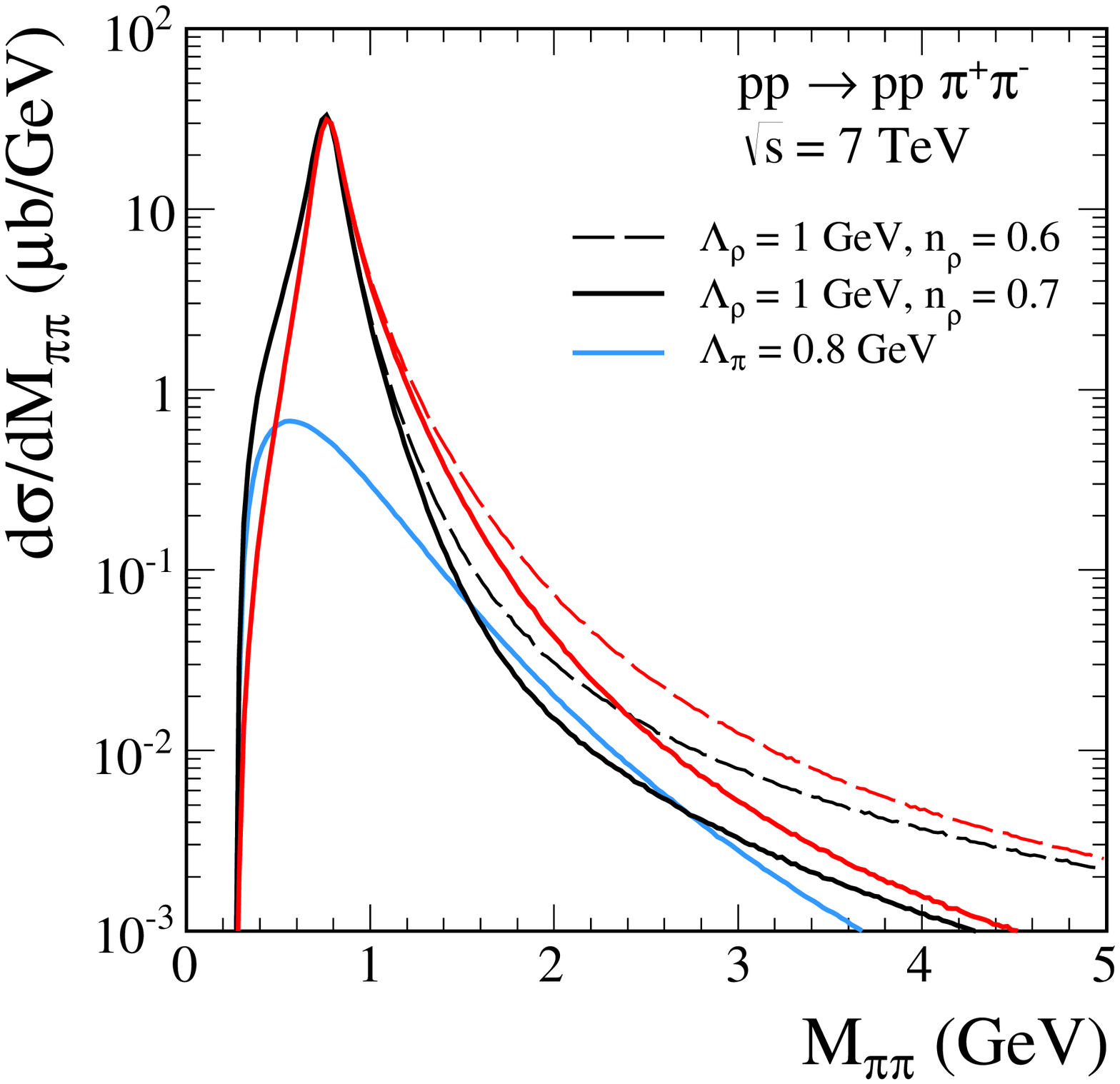}
\caption{\label{fig:rho_pipi_M34}
Two-pion invariant mass distributions at $\sqrt{s}=7$~TeV for two different scenarios discussed in the main text.
The $\rho^{0}$ (red lines), continuum (blue lines) and total (black lines)
contributions are shown.
The left panel corresponds to a ``hard'' form factor (\ref{ff_Nachtmann})
and the form factor $F(p_{t}^{2},p_{u}^{2},p_{34}^{2}) = 1$ in (\ref{ff_Poppe})
for the inner processes.
The right panel corresponds to a ``soft'' form factor (\ref{ff_Nachtmann})
and a non-trivial form factor (\ref{monopole_ff}) for the inner processes.
%The solid line corresponds to the pomeron and $f_{2 \Reg}$ exchanges in the amplitude
%while the long-dashed (blue) line corresponds to the pomeron exchange alone.
%The black line represents parameters set~A of coupling constants given by (\ref{setA}) 
%both for the $\Gamma^{(0)}$ and $\Gamma^{(2)}$ tensor functions,
%while the blue line represents set~B (\ref{setB}) for the $\Gamma^{(2)}$ function alone.
}
\end{figure}

%We calculate the non-resonant (Drell-S\"oding) contribution and discuss 
%its interference effects with the resonant contribution.
Now we turn to the full calculation including
the resonant and the non-resonant (Drell-S\"oding) contributions.
These have to be added at the amplitude level and then squared
to get the cross sections including all interference effects.
In Fig.~\ref{fig:rho_pipi_M34} we show the resulting
two-pion invariant-mass distribution for two different scenarios.
In the first scenario (left panel) we take a relatively hard form factor
for the resonant contribution and no form factors
for the inner process for the non-resonant contribution.
In our calculations we set $\Lambda_{\rho} = 2$~GeV and 
$n_{\rho} = 0.4, 0.5$, or 0.6 in (\ref{ff_Nachtmann}) and 
$F(p_{t}^{2},p_{u}^{2},p_{34}^{2}) = 1$ in (\ref{ff_Poppe}).
This is the scenario discussed recently in \cite{Bolz:2014mya}.
In this scenario strong interference effects can be observed.
These effects strongly depend on the details of the form factor
(compare the three different lines, 
corresponding to the three different values of $n_{\rho}$, 
in the left panel of Fig.~\ref{fig:rho_pipi_M34}).
In the present paper we consider also a second scenario (right panel) 
with softer form factor (\ref{ff_Nachtmann}) 
($\Lambda_{\rho} = 1$~GeV, $n_{\rho}$ = 0.6 and 0.7)
and including monopole-like form factors (\ref{ff_Poppe}), (\ref{monopole_ff}),
for the inner processes with $\Lambda_{\pi} = 0.8$~GeV.
The final result for the two scenarios is rather similar.
Therefore, in the following presentation we shall use only the first scenario
as representative.
%--------------------------------------------------------
%--------------------------------------------------------

\begin{figure}[tbp]
\centering
\includegraphics[width=0.48\textwidth]{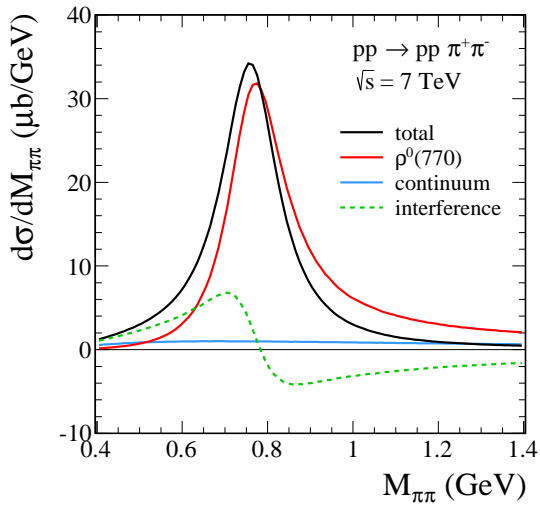}
\caption{\label{fig:rho_pipi_M34_interference}
Two-pion invariant mass distribution at $\sqrt{s}=7$~TeV
for the resonant and non-resonant contributions as well as their coherent sum.
The interference term of $\rho^{0}$ with the $\pi^{+} \pi^{-}$ continuum is also shown.
We have taken here $\Lambda_{\rho} = 2$~GeV and $n_{\rho} = 0.5$ in (\ref{ff_Nachtmann}).
}
\end{figure}

In Fig.~\ref{fig:rho_pipi_M34_interference} we show the interference effect
in the $\rho^{0}$ mass region.
The skewing of the $\rho^{0}$ shape caused by the interference of 
the resonant and non-resonant contributions is clearly exposed.
The skewing depends crucially on the choice of the $\rho^{0}$ form 
factor parameterisation (\ref{ff_Nachtmann}).
%--------------------------------------------------------
%--------------------------------------------------------

\begin{figure}[tbp]
\centering
\includegraphics[width=0.48\textwidth]{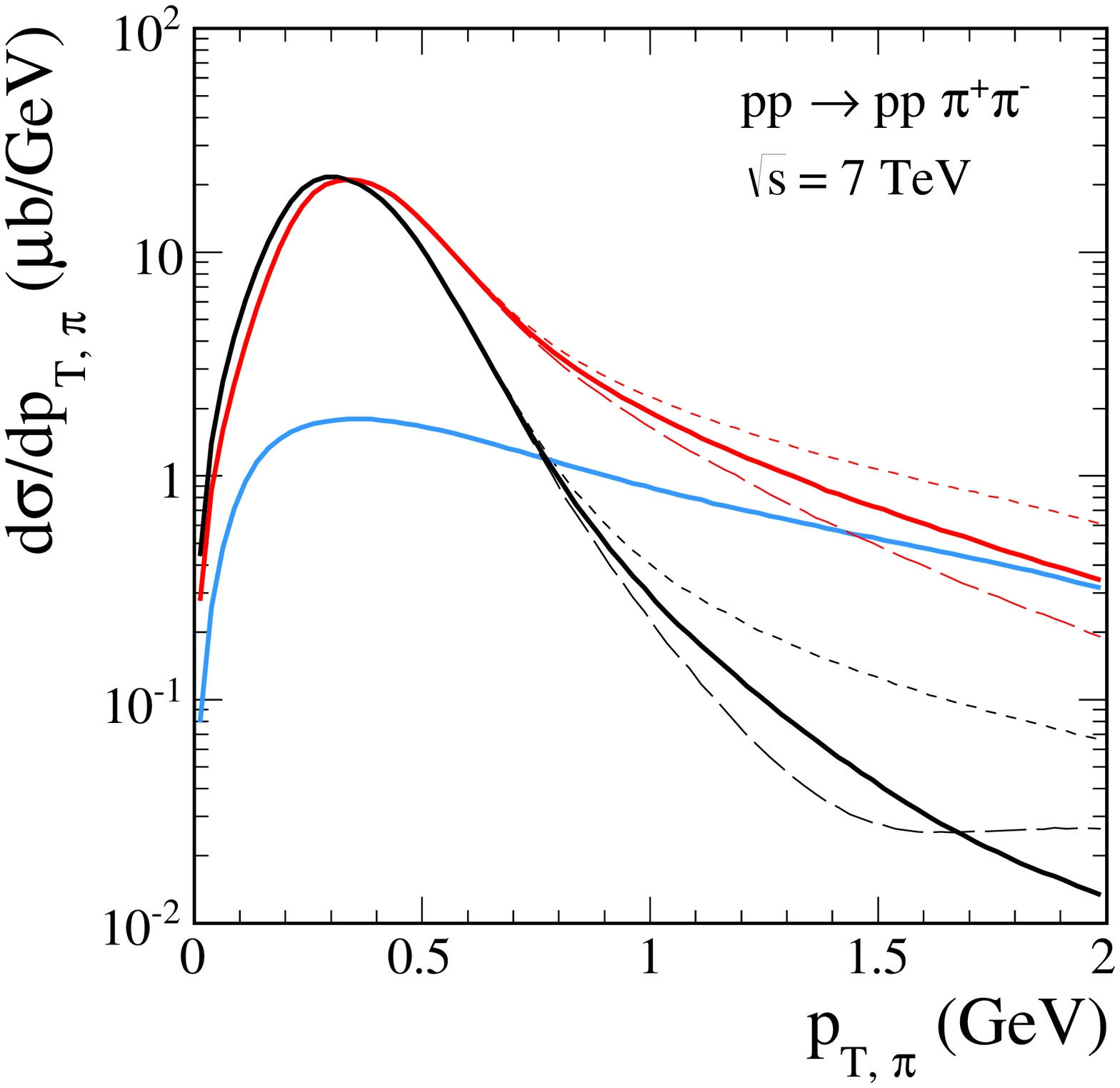}
\caption{\label{fig:rho_pipi_p3t}
The distributions in pion transverse momentum
for the central exclusive $\rho^{0} \to \pi^{+}\pi^{-}$ and continuum $\pi^{+} \pi^{-}$ production
at $\sqrt{s} = 7$~TeV. The meaning of the lines
is the same as in Fig.~\ref{fig:rho_pipi_M34} (left panel).
%The solid line corresponds to the pomeron and $f_{2 \Reg}$ exchanges in the amplitude
%while the long-dashed (blue) line corresponds to the pomeron exchange alone.
}
\end{figure}

In Fig.\ref{fig:rho_pipi_p3t} we present distributions
in pion transverse momentum $p_{\perp,\pi}$,
i.e. $p_{\perp,\pi^{+}}$ or $p_{\perp,\pi^{-}}$.
%The general behaviour reminds that for the dipion invariant mass distribution.
At small $p_{\perp,\pi}$ the resonance contribution
is the dominant one. At higher $p_{\perp,\pi}$ our calculation,
with the chosen model parameters, gives a strong cancellation
between the resonant and the non-resonant terms.
%--------------------------------------------------------
%--------------------------------------------------------

%In Fig.~\ref{fig:rho_pipi_xi1xi2} we show corresponding two-dimensional
%distributions in $(\xi_1,\xi_2)$ in the full phase space.
%--------------------------------------------------------
%\begin{figure}[!ht]
%\includegraphics[width=0.48\textwidth]{map_xi1xi2_200.eps}
%\includegraphics[width=0.48\textwidth]{map_xi1xi2_7000.eps}
%  \caption{\label{fig:rho_pipi_xi1xi2}
%  \small
%Distribution in  $(\xi_1,\xi_2)$ = 
%$(\log_{10}(p_{1\perp}/1\,\mathrm{GeV}), 
%  \log_{10}(p_{2\perp}/1\,\mathrm{GeV}))$
%for the $\rho^{0} \to \pi^{+}\pi^{-}$ contribution
%at $\sqrt{s} = 200$~GeV (left panel) and $7$~TeV (right panel).
%}
%\end{figure}
%--------------------------------------------------------

%--------------------------------------------------------
%\begin{figure}[!ht]
%\includegraphics[width=0.48\textwidth]{dsig_dpt1_200.eps}
%\includegraphics[width=0.48\textwidth]{dsig_dpt1_1960.eps}
%\includegraphics[width=0.48\textwidth]{dsig_dpt1_7000.eps}
%  \caption{\label{fig:rho_pipi_y}
%  \small
%}
%\end{figure}
%--------------------------------------------------------

\begin{figure}[tbp]
\centering
\includegraphics[width=0.48\textwidth]{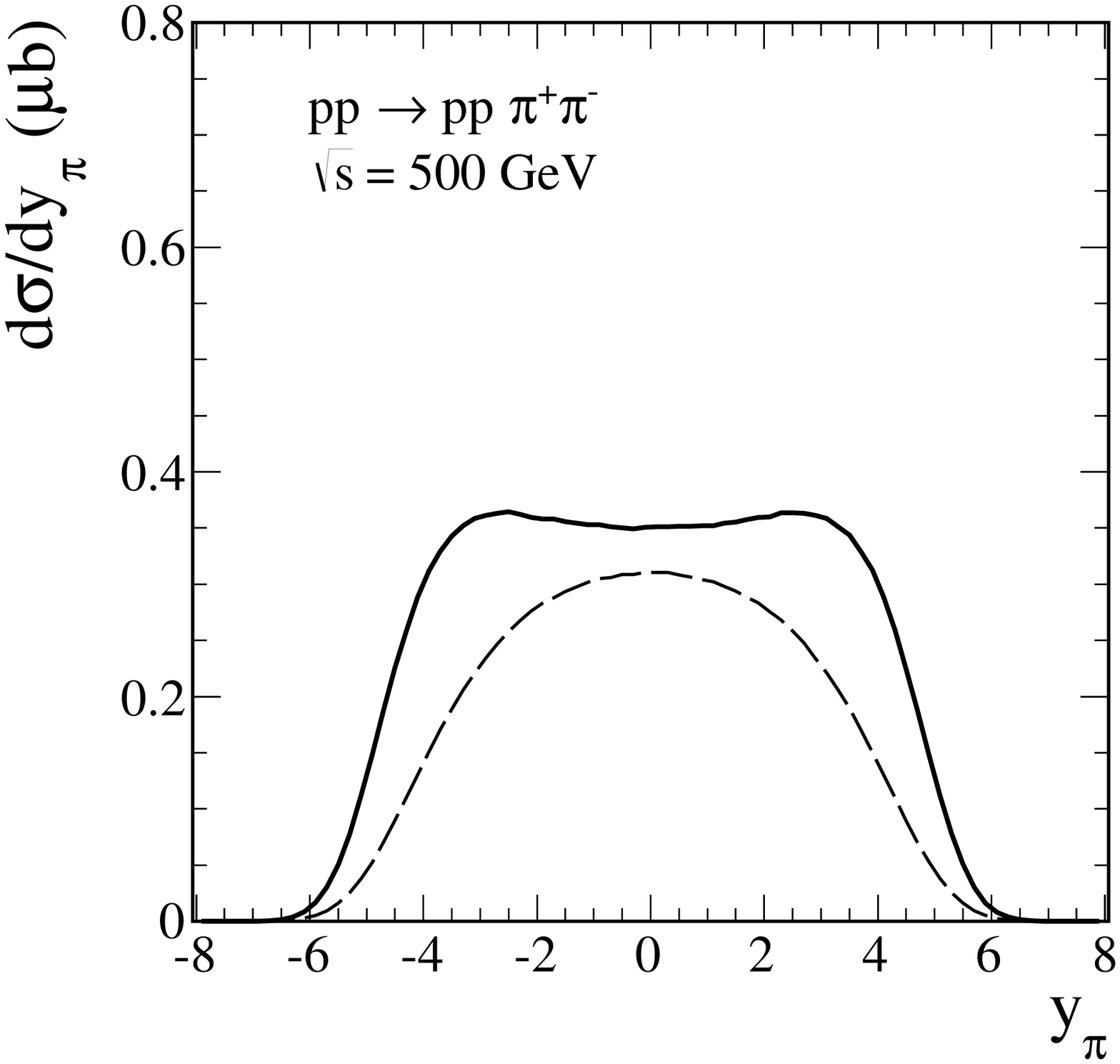}
\includegraphics[width=0.48\textwidth]{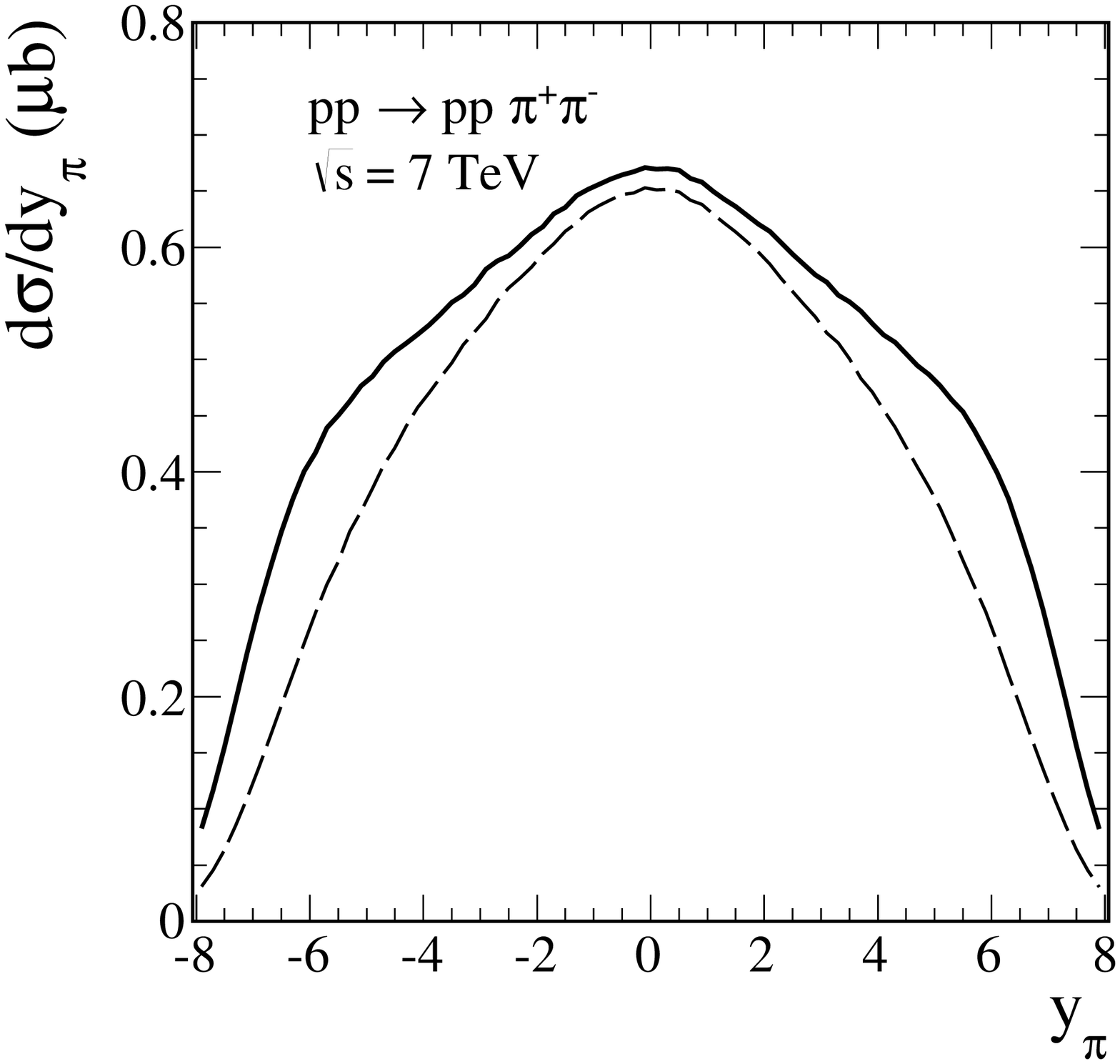}\\
\includegraphics[width=0.48\textwidth]{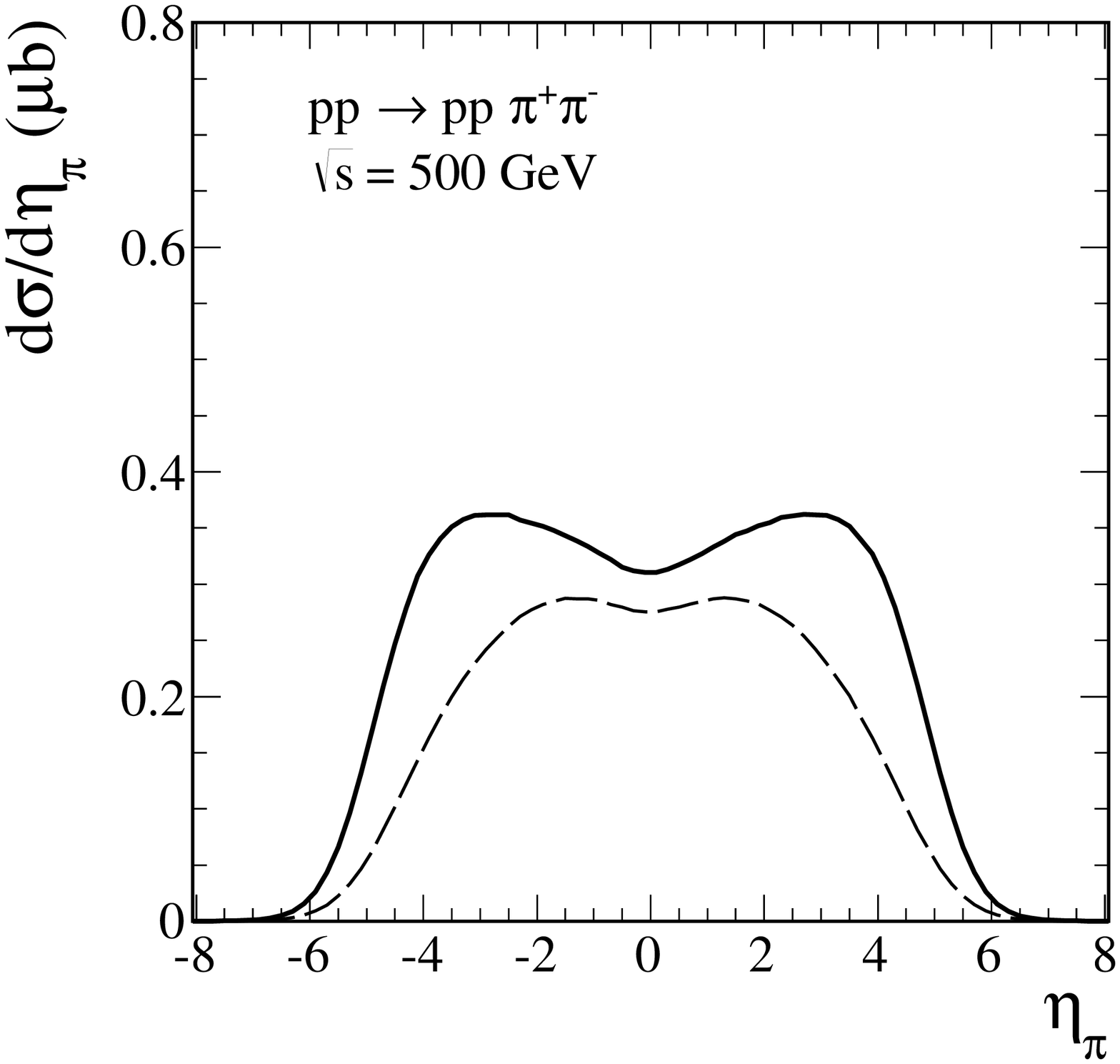}
\includegraphics[width=0.48\textwidth]{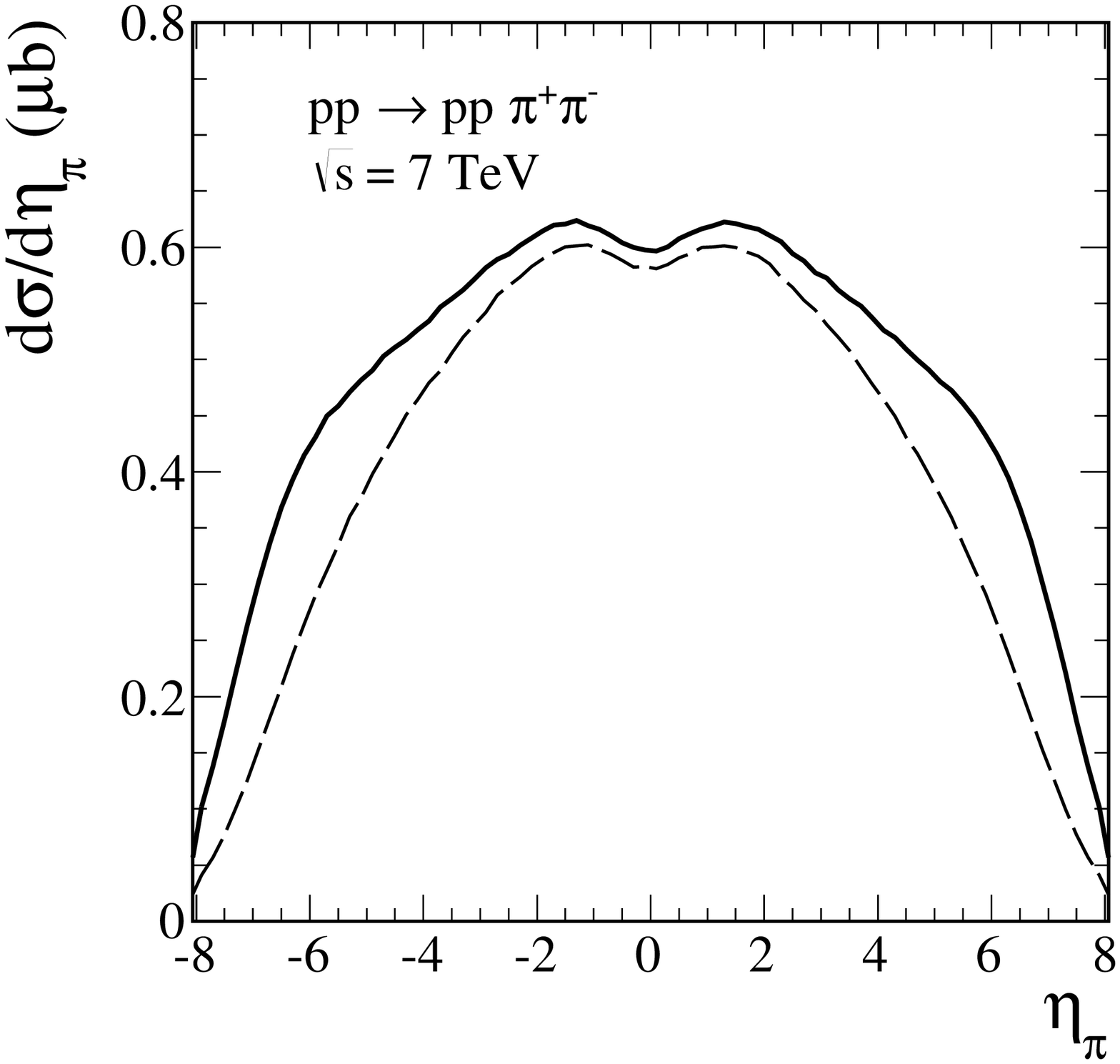}
\caption{\label{fig:rho_pipi_y}
Rapidity (upper row) and pseudorapidity (lower row) distributions 
of pions produced in the photoproduction mechanism
at $\sqrt{s} = 0.5$ and $7$~TeV.
The solid lines correspond to the pomeron and $f_{2 \Reg}$ exchanges in the amplitudes
while the dashed lines correspond to the pomeron exchange alone.}
\end{figure}

In Fig.~\ref{fig:rho_pipi_y} we show rapidity $\mathrm{y}_{\pi}$ 
and pseudorapidity $\eta_{\pi}$ distributions
of pions produced in the photoproduction mechanism.
The $f_{2 \Reg}$ exchange included in the amplitude 
contributes mainly at backward and forward pion rapidities.
Its contribution is nonnegligible even at the LHC.
The dip in the $\eta_{\pi}$ distribution for $|\eta_{\pi}| \to 0$
is a kinematic effect; see Appendix~D of \cite{Lebiedowicz:2013ika}.
%--------------------------------------------------------
%--------------------------------------------------------

\begin{figure}[tbp]
\centering
\includegraphics[width=0.5\textwidth]{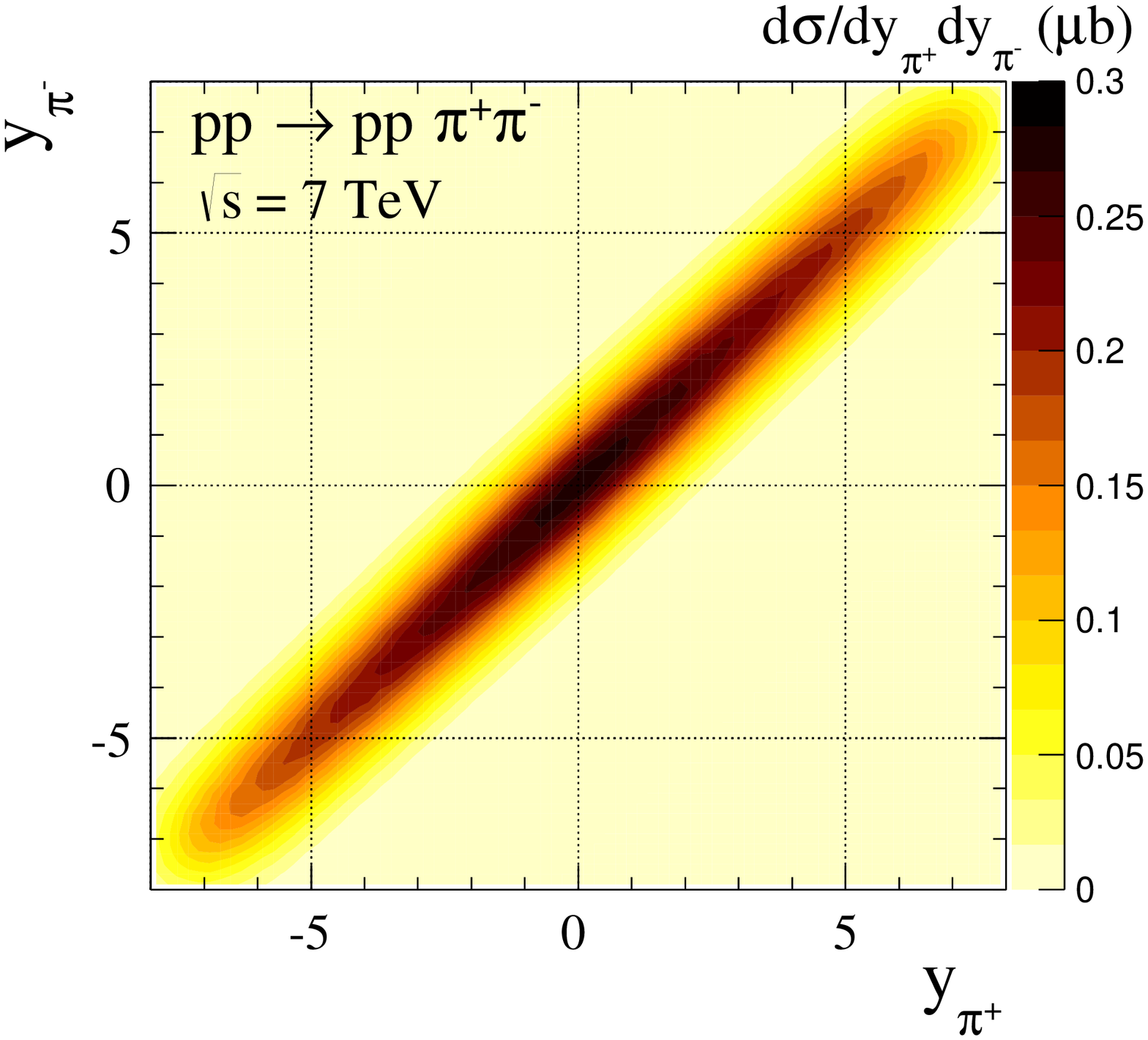}
\caption{\label{fig:map_y3y4}
%Differential cross section in $(y_3,y_4)$ space
The correlation between the rapidities $\mathrm{y}_{\pi^+}$ and $\mathrm{y}_{\pi^-}$
for the photoproduction mechanism at $\sqrt{s} = 7$~TeV.
Plotted is $d\sigma/d\mathrm{y}_{\pi^{+}}d\mathrm{y}_{\pi^{-}}$ in $\mu$b.
%at $\sqrt{s} = 200$~GeV (left panel) and $7$~TeV (right panel).
%In this calculation we have used the parameter set~B of coupling 
%constants given by (\ref{setB}).
}
\end{figure}

The correlation between $\mathrm{y}_{\pi^+}$ and $\mathrm{y}_{\pi^-}$ is
displayed in Fig.~\ref{fig:map_y3y4}. 
One can observe a strong correlation between $\mathrm{y}_{\pi^+}$ and $\mathrm{y}_{\pi^-}$.
%and symmetry. Compare, e.g., 
This correlation is similar to the correlation studied 
in \cite{Lebiedowicz:2009pj,Lebiedowicz:2011nb}
for the Regge-like diffractive production of $\pi^+ \pi^-$ continuum,
i.e. the pions are emitted preferentially in the same hemispheres.
In contrast to the $\Pom/\Reg$-$\Pom/\Reg$ fusion \cite{Lebiedowicz:2009pj},
where the camel-like shape was predicted,
no large $|\mathrm{y}_{\pi}|$ enhancement is predicted for the photoproduction case.
%--------------------------------------------------------
%--------------------------------------------------------

\begin{figure}[tbp]
\centering
\includegraphics[width=0.48\textwidth]{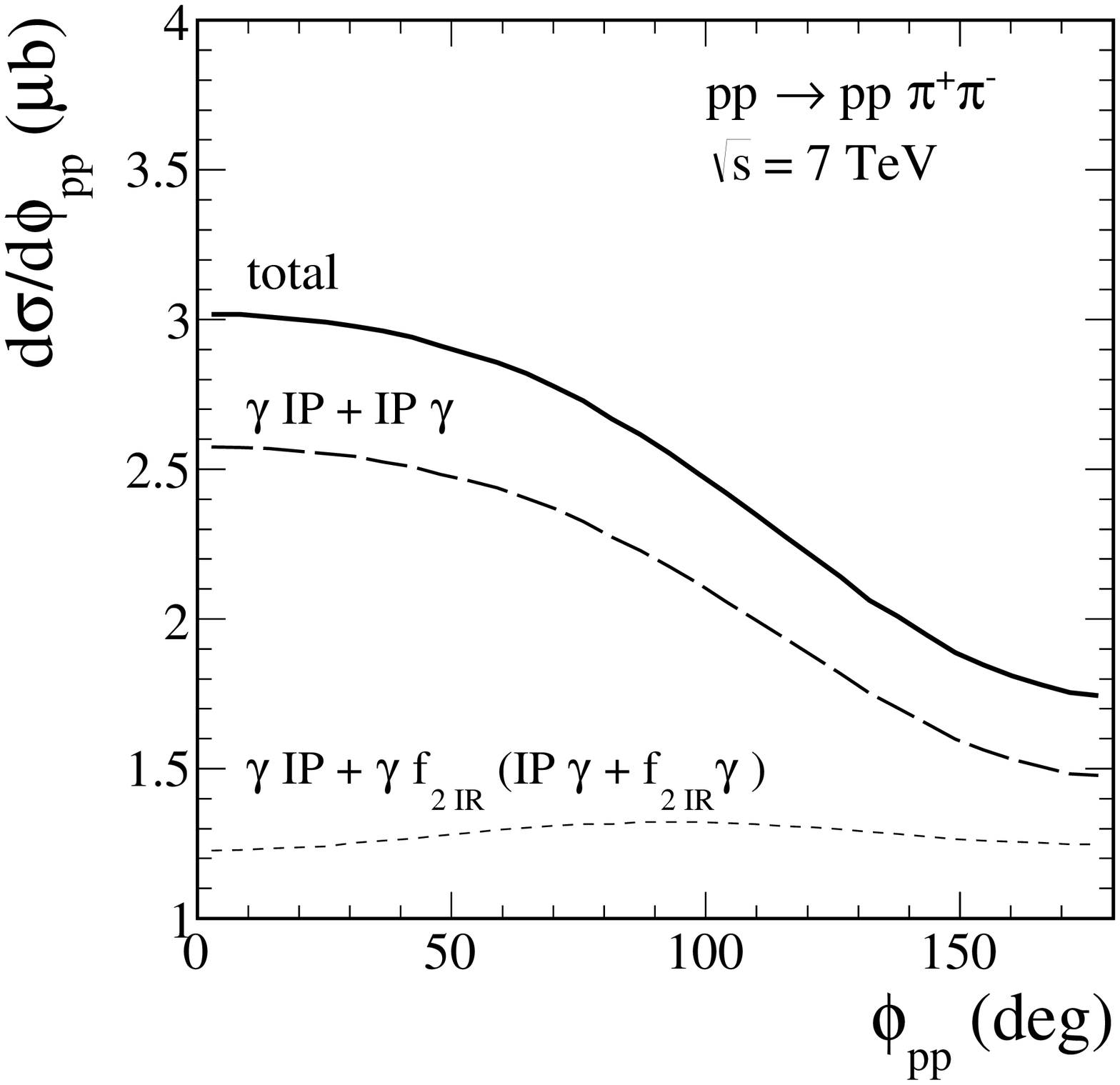}
\includegraphics[width=0.48\textwidth]{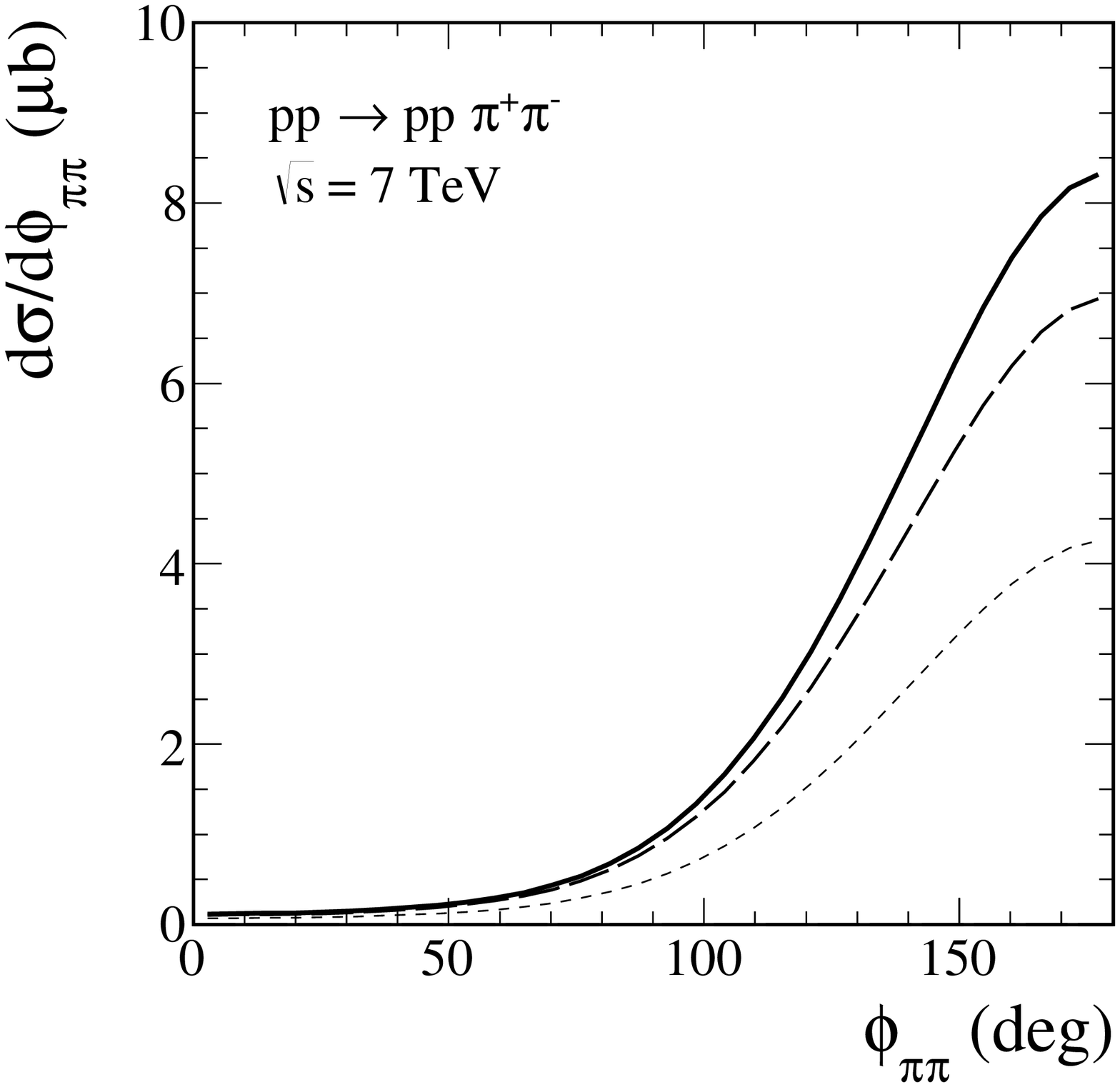}
\caption{\label{fig:rho_pipi_phi}
The distribution in azimuthal angle between the outgoing protons (left panel)
and between the outgoing pions (right panel)
for the central exclusive photoproduction at $\sqrt{s} = 7$~TeV.
The solid lines correspond to the pomeron and $f_{2 \Reg}$ exchanges in the amplitudes
while the long-dashed lines correspond to the pomeron exchange alone.
The short-dashed line represents the contribution
from $\gamma-\Pom/\Reg$ (or $\Pom/\Reg-\gamma$) exchange.
%The lines are described in Fig.~\ref{fig:rho_pipi_xi1}.
}
\end{figure}

In Fig.~\ref{fig:rho_pipi_phi} we show the cross section 
as a function of the azimuthal angle between 
the transverse momentum vectors of the two outgoing protons (left panel)
and between the outgoing pions (right panel)
for the photoproduction mechanism.
In the first case the effect of deviation from a constant is due to
interference of photon-pomeron and pomeron-photon amplitudes.
The interference is different for $pp$- and $p \bar{p}$-collisions
because proton and antiproton have opposite charges.
A similar effect was first discussed in Ref.~\cite{Schafer:2007mm} 
for the exclusive production of $J/\psi$ mesons. 
These correlations are rather different than those for the double-pomeron mechanism 
\cite{Lebiedowicz:2011nb,Lebiedowicz:thesis} and could therefore be
used for at least a partial separation of the two mechanisms.
%The bremsstrahlung contribution calculated 
%in the Born approximation
%by Eq.~(\ref{amplitude_bremsstrahlung_c}) peaks at $\phi_{pp} = \pi$.
The absorption effects, see e.g.
\cite{Cisek:2011vt, Lebiedowicz:thesis},
lead to extra decorrelation in azimuth compared to the Born-level results presented here. 
%--------------------------------------------------------
%--------------------------------------------------------

%--------------------------------------------------------
%\begin{figure}[!ht]
%\includegraphics[width=0.48\textwidth]{dsig_dphi12_200_bremsstr.eps}
%\includegraphics[width=0.48\textwidth]{dsig_dphi12_1960_bremsstr.eps}
%\includegraphics[width=0.48\textwidth]{dsig_dphi12_7000_bremsstr.eps}
%  \caption{\label{fig:rho_pipi_phi_bremsstr}
%  \small
%Distributions in azimuthal angle between outgoing protons
%for the photoproduction $\rho^{0}$-resonance contribution and
%the contribution given by Eq.~(\ref{amplitude_bremsstrahlung_c}) 
%at $\sqrt{s} = 0.2, 1.96$ and $7$~TeV.
%The line description is the same as in Fig.~\ref{fig:rho_pipi_xi1}.
%}
%\end{figure}
%--------------------------------------------------------

%Two-pion invariant masses are shown in Fig.~\ref{fig:rho_pipi_M34}
%for three different center-of-mass energies.

\begin{figure}[tbp]
\centering
\includegraphics[width=0.48\textwidth]{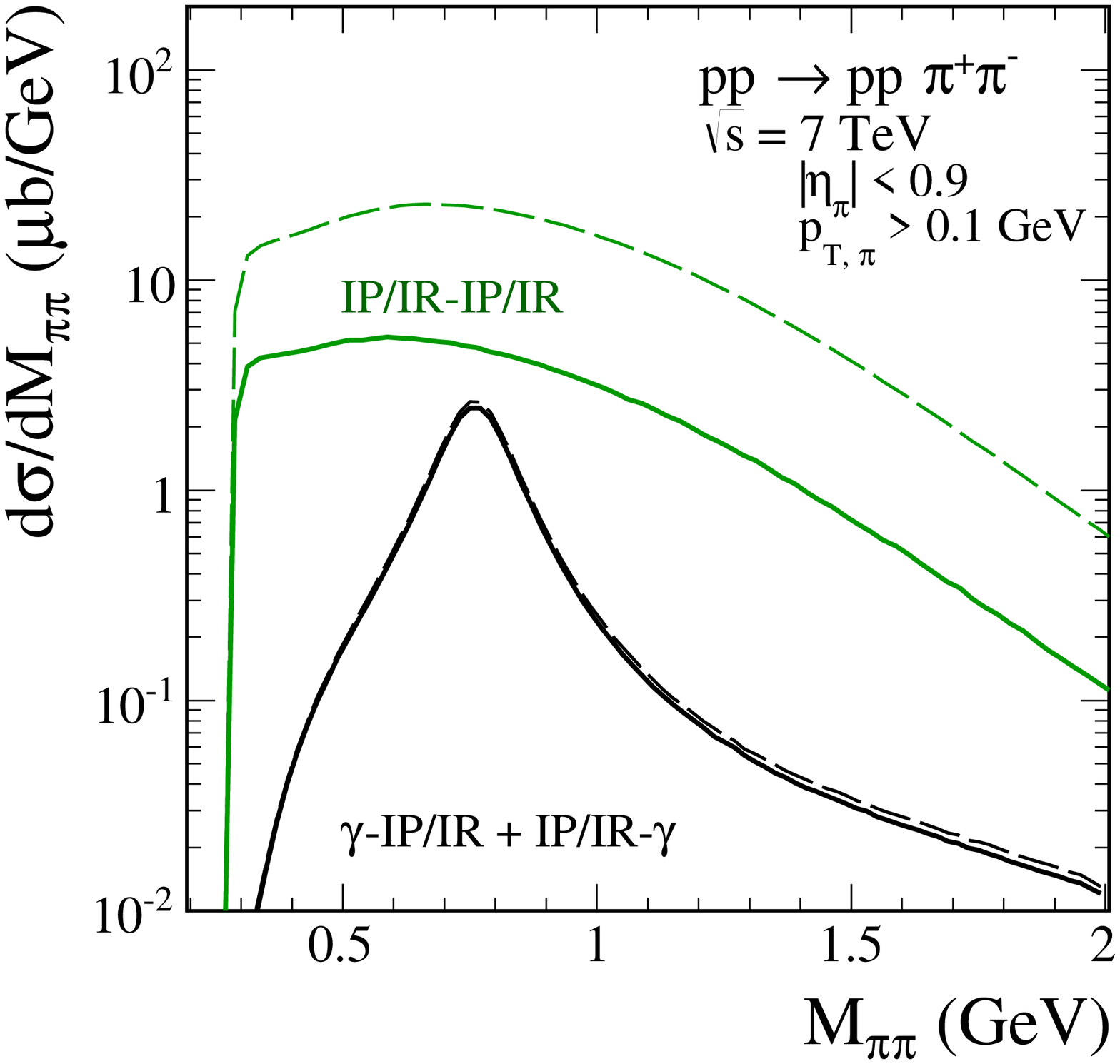}
\includegraphics[width=0.48\textwidth]{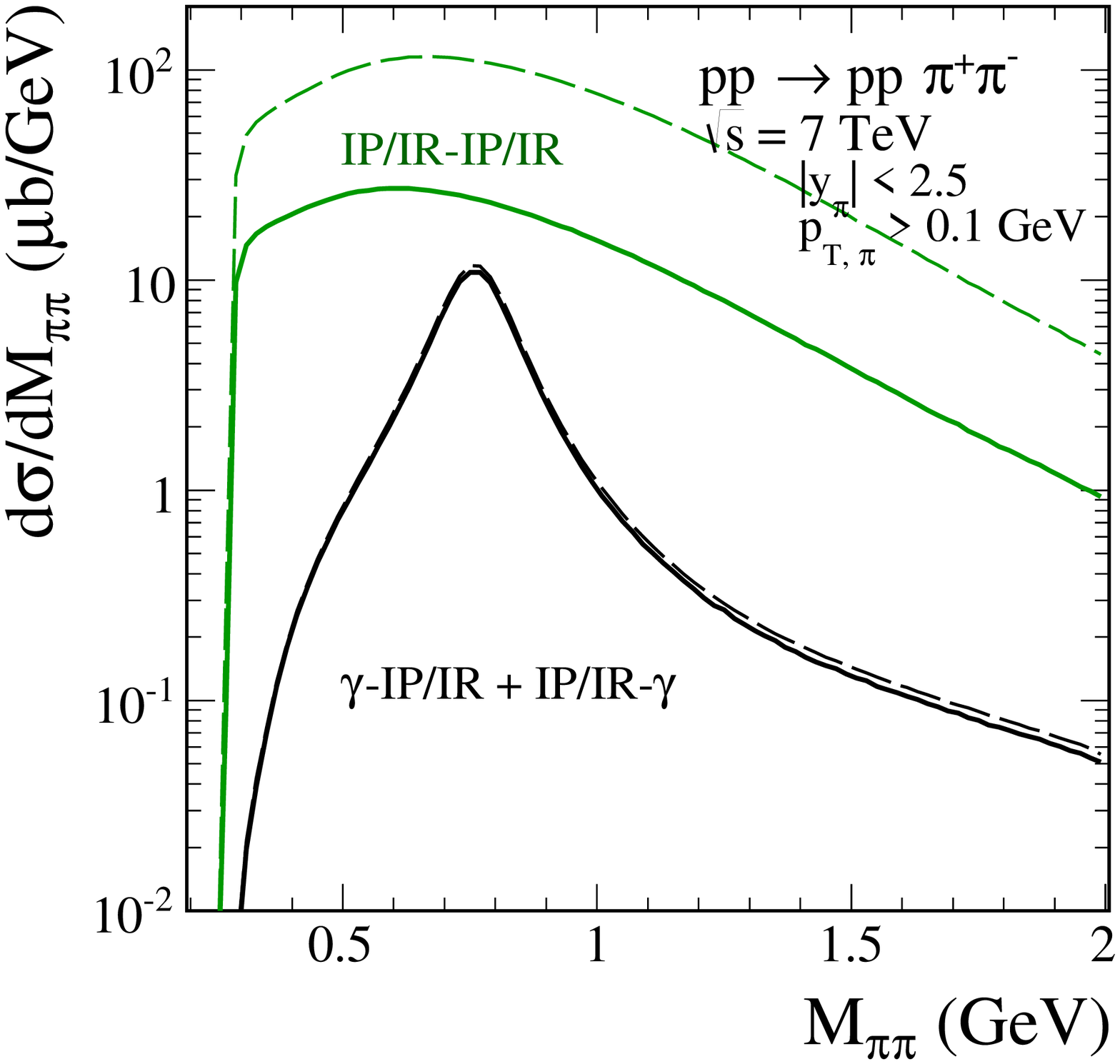}
\caption{\label{fig:two_mechanisms}
Two-pion invariant mass distributions at $\sqrt{s}=7$~TeV
with kinematical cuts specified in the figure.
We show results for the double-pomeron/reggeon contribution
calculated in the simple Lebiedowicz-Szczurek Regge-like model 
\cite{Lebiedowicz:2009pj,Lebiedowicz:thesis}
and from the photon-pomeron/reggeon contribution as discussed in 
the present paper, without (dashed lines) and with (solid lines) 
absorption effects due to the $pp$-interaction.
The calculation of double-pomeron/reggeon contribution 
was done with the cut-off parameter $\Lambda_{off,E}^{2} = 1.2$~GeV$^{2}$
in the off-shell pion form factors 
(see Section~2.3 of \cite{Lebiedowicz:thesis}).
%The dotted lines show the respective ratios
%$d\sigma^{Born\,+\,pp-rescattering}/d\sigma^{Born}$.
%for two different forms of off-shell pion form factor given by the black and blue lines
%and for the $\rho^{0}$ contribution given by the green line.
%Right: Invariant mass distribution of two particles, assumed to be $\pi^{+} \pi^{-}$
%produced in proton-antiproton collisions at $\sqrt{s}=1.96$~TeV
%measured by the CDF detector at the Fermilab Tevatron.
}
\end{figure}

Finally, in Fig.~\ref{fig:two_mechanisms} 
we compare the photoproduction contribution 
with the double-pomeron/reggeon contribution one
(for details see \cite{Lebiedowicz:2009pj,Lebiedowicz:2011nb,Lebiedowicz:2011tp}).
The absorption effects due to $pp$-interaction lead to huge damping
of the cross section for pomeron-pomeron fusion and relatively small
reduction of the cross section for the photoproduction mechanism.
The dependence of absorption on $M_{\pi\pi}$ is quantified
by the ratio of full and Born cross sections
$<S^{2}(M_{\pi\pi})> = 
\frac{d\sigma^{Born\,+\,pp-rescattering}/dM_{\pi\pi}}{d\sigma^{Born}/dM_{\pi\pi}}$.
We obtain $<S^{2}> \simeq 0.9$ for the photon-pomeron/reggeon contribution
and $<S^{2}> \simeq 0.2$ for the double-pomeron/reggeon contribution.
We observe that at midrapidities,
imposing e.g. the cut $|\eta_{\pi}|<0.9$,
the photoproduction term could be visible in experiments.
%The absorptive corrections will be discussed in a forthcoming paper \cite{LS_absorption}.
%--------------------------------------------------------
%--------------------------------------------------------

In the present study we do not include interference effects between the
photoproduction and the purely diffractive amplitudes.
This goes beyond the scope of the present paper.
We leave a detailed analysis of such effects for future studies
when the diffractive mechanism with tensorial pomeron/reggeon will be discussed.
%However, we do not expect spectacular effects.

In Table \ref{tab:table} we have collected cross sections in $\mu b$
for the exclusive $\pi^{+}\pi^{-}$ photoproduction
contributions, see Figs.~\ref{fig:gampom_pomgam_s} and 
\ref{fig:gampom_pomgam_b}, 
for kinematical range of ``$\rho^{0}$ - mass window''
$2 m_{\pi} \leqslant M_{\pi\pi} \leqslant 1.5$~GeV
and when some experimental cuts are imposed.
%=======================================================================================
\begin{table}[tbp]
\centering
\begin{tabular}{|c|c|c|}
\hline
$\sqrt{s}$, TeV & $\Pom$ and $f_{2 \Reg}$  & $\Pom$ \\
\hline
0.2         & \;\;2.22  & \;\;1.36 \\
0.5         & \;\;3.38  & \;\;2.31 \\
1.96        & \;\;5.77  & \;\;4.51 \\
7           & \;\;7.80  & \;\;6.64 \\
%\hline
%7 ($\rho^{0}$ + continuum)           &  12.00     &  \\
%\hline
%7 (continuum only)           &  4.85     &  \\
\hline
%0.2& STAR-I cuts   &  0.032 (0.038)  & 0.026 \\ 
%0.5& STAR-II cuts  &  0.004 (0.004)  & 0.004\\ 
0.2 (STAR cuts)     & \;\;0.011  & \;\;0.008 \\ 
7 (ALICE cuts)      & \;\;0.59  & \;\;0.58 \\ 
7 (CMS cuts)        & \;\;2.61  & \;\;2.53 \\
\hline
\end{tabular}
\caption{\label{tab:table} The integrated cross sections in $\mu b$ 
for the central exclusive $\pi^{+}\pi^{-}$ production
in $pp$ collisions via the photoproduction mechanism
without and with some typical experimental cuts.
The line with $\sqrt{s} = 1.96$~TeV corresponds to the $p \bar{p}$ collision.
The column ``$\Pom$ and $f_{2 \Reg}$'' 
shows the resulting total cross sections from $\Pom$ and $f_{2 \Reg}$ exchanges,
which include, of course, the interference term between the two components.
The column ``$\Pom$'' shows results obtained for the $\Pom$ exchange alone.
In the calculations for the last three lines the following cuts were imposed:
%for the STAR detector at RHIC
$|\eta_{\pi}| < 1.0$, $|\eta_{\pi \pi}| < 2.0$, $p_{\perp, \pi} > 0.15$~GeV,
0.005~$< -t_{1}, -t_{2} < 0.03$~GeV$^{2}$ at $\sqrt{s} = 200$~GeV (STAR cuts)
while at the LHC energies
$|\mathrm{y}_{\pi}| < 2.5$, $p_{\perp, \pi} > 0.1$~GeV (CMS cuts) and
$|\eta_{\pi}| < 0.9$, $p_{\perp, \pi} > 0.1$~GeV (ALICE cuts).}
\end{table}
%========================================================================================

%-------------------------------------------------------------------
\section{Conclusions and Outlook}
\label{sec:section_5}
%-------------------------------------------------------------------

In the present paper we have made first estimates of the $\rho^{0}$ resonance 
and the Drell-S\"oding contributions to 
the $p p \to p p \pi^+ \pi^-$ and $p \bar{p} \to p \bar{p} \pi^+ \pi^-$ reactions.
We have shown several differential distributions in pion rapidities
and transverse momenta as well as some observables related to 
final state protons. 
The photoproduction contribution constitutes several percent
of the double-pomeron/reggeon contribution
presented here in a simple Regge-like model, strongly depending on 
the invariant mass of the two-pion system.
In the photoproduction contribution 
%has similar characteristics 
%as far as rapidity distributions are considered.
the rapidities of the two pions are strongly correlated
and $\mathrm{y}_{\pi^+} \approx \mathrm{y}_{\pi^-}$. This is similar as for
the double-pomeron/reggeon exchanges in the fully diffractive mechanism.
The dependence on transverse momenta of pions is somewhat different
for the two contributions.
Therefore, including transverse momentum dependent cuts 
changes the relative size of the photoproduction 
and double-pomeron/reggeon contributions. 

Our cross section for $p p \to p p (\rho^0 \to \pi^+ \pi^-)$
is slightly larger than for the $p p \to p p \rho^0$ with fixed sharp 
resonance mass. This is partly due to the different phase space for 
the four-body and three-body reactions.

The $\rho^0$-photoproduction and purely diffractive 
contributions have different dependences on the proton transverse momenta. 
This could be used to separate the $\rho^0$ contribution. 
Also azimuthal angle correlations between protons are quite different.
One could separate the space in azimuthal angle into two regions:
$\phi_{pp} < 90^o$ and $\phi_{pp} > 90^o$. 
The contribution of $\rho^0$ in the first region should be strongly enhanced
for $pp$-collisions (this is not true for $p \bar{p}$-collisions).
We therefore conclude that the measurement of forward/backward
protons is crucial for a better understanding of the mechanism
of the $p p \to p p \pi^+ \pi^-$ reaction, see e.g. \cite{Staszewski:2011bg}. 
Both, the ALFA detector associated with the ATLAS main detector 
and particularly the TOTEM detector associated with the CMS main detector 
could be used to measure protons and pions in coincidence, see \cite{Osterberg:2014mta}.
Also a cut on $\phi_{\pi \pi}$ could help to enhance the $\rho^{0}$ contribution.

In general, the $\rho^{0}$ and the whole photoproduction contribution
could be obtained from partial wave analysis of the experimental results.
Whether this is possible requires dedicated Monte Carlo studies.

We have also discussed the role of soft $pp$-rescattering corrections
which lead to a shape deformation of differential distributions 
in contrast to the commonly used uniform factor known as gap survival factor.
%In the present paper we have neglected absorption effects which 
The absorptive corrections for photon induced reactions
lead to about 10\% reduction of the cross section.
In the case of the photon-mediated contribution
to central exclusive production of heavier mesons 
the effect is even larger, see e.g. \cite{Cisek:2014ala}.
%For the four-body reaction discussed here a similar effect is expected, 
%i.e. a large, energy-dependent, damping of the cross section
%and a modification of the distributions, in particular, for the
%transverse momenta of the outgoing protons;
%see e.g. results in Section~2.6 of \cite{Lebiedowicz:thesis}

In future work we shall present a calculation of the remaining processes
leading to central exclusive $\pi^{+} \pi^{-}$ production
in $pp$ and $p \bar{p}$ collisions.
These are, as already mentioned in the introduction,
continuum $\pi^{+} \pi^{-}$ production via pomeron/reggeon - pomeron/reggeon
fusion and $f_{2}(1270)$ production by the same mechanisms with the
$f_{2}$ decaying into $\pi^{+} \pi^{-}$.
All these reactions can be treated in the model for high-energy
soft reactions presented in \cite{Ewerz:2013kda}.
Concerning the photon-mediated processes discussed in the present paper
one could also include the Pauli form factor
in the proton's electromagnetic vertex as was done recently 
for exclusive production of $J/\psi$ meson \cite{Cisek:2014ala}.
%The tensor coupling is important for large $|t_{1}|$ or $|t_{2}|$ and as a consequence
The effect of enhancements is expected only at large transverse momenta 
of the dipion pairs.

To summarise: using the tensor-pomeron approach
we have treated the photon-mediated contributions
to the central exclusive production of $\pi^{+}\pi^{-}$
in $pp$ and $p \bar{p}$ collisions.
The couplings involved are, in essence, known.
Thus the predictions made with our model should be quite solid.
This should give guidelines and a target to shoot at for experimentalists
working on soft high-energy reactions at colliders.

\acknowledgments

%The authors would like to thank ..
We are indebted to C. Ewerz, M. Sauter, W. Sch\"afer, R. Schicker,
and A. Sch\"oning for useful discussions.
The work of P.L. was supported
by the Polish NCN grant DEC-2013/08/T/ST2/00165 (ETIUDA) 
%National Science Centre on the basis of decision
and by the MNiSW grant no. IP2014~025173 ``Iuventus Plus''
%by Polish Ministry of Science and Higher Education
as well as by the START fellowship from the Foundation for Polish Science.
The work of A.S. was partially supported by the Polish NCN grant
DEC-2011/01/B/ST2/04535 (OPUS) as well as by the Centre for Innovation and
Transfer of Natural Sciences and Engineering Knowledge in Rzesz\'ow.

%\paragraph{Note added.} This is also a good position for notes added
%after the paper has been written.

%------------------------------------------------------------------
%%% bibliography
\nocite{}

{
\begin{small}
\bibliography{refs}

\providecommand{\href}[2]{#2}\begingroup\raggedright\begin{thebibliography}{10}

\bibitem{Austregesilo:2013yxa}
{\bf COMPASS} Collaboration, A.~Austregesilo, {\it {A Partial-Wave Analysis of
  Centrally Produced Two-Pseudoscalar Final States in $pp$ Reactions at
  COMPASS}},  {\em PoS} {\bf Bormio2013} (2013) 014.

\bibitem{Turnau_DIS2014}
{\bf STAR} Collaboration, J.~Turnau, {\it {Measurement of the Central Exclusive
  Production of pion pairs using tagged forward protons at the STAR detector at
  RHIC}},  {\em PoS} {\bf DIS2014} (2014) 098.

\bibitem{Adamczyk:2014ofa}
L.~Adamczyk, W.~Guryn, and J.~Turnau, {\it {Central exclusive production at
  RHIC}},  \href{http://arxiv.org/abs/1410.5752}{{\tt arXiv:1410.5752}}.

\bibitem{Albrow:2013mva}
{\bf CDF} Collaboration, M.~Albrow, A.~{\'S}wi\c{e}ch, and M.~{\.Z}urek, {\it
  {Exclusive Central $\pi^+ \pi^-$ production in CDF}},
  \href{http://arxiv.org/abs/1310.3839}{{\tt arXiv:1310.3839}}.

\bibitem{Albrow_Project_new}
M.~Albrow, J.~Lewis, M.~{\.Z}urek, A.~{\'S}wi\c{e}ch, D.~Lontkovskyi,
  I.~Makarenko, and J.~S. Wilson. The public note called \textit{Measurement of
  Central Exclusive Hadron Pair Production in CDF} is available at
  \url{http://www-cdf.fnal.gov/physics/new/qcd/GXG_14/webpage/}.

\bibitem{Schicker:2012nn}
{\bf ALICE} Collaboration, R.~Schicker, {\it {Central Diffraction in ALICE}},
  \href{http://arxiv.org/abs/1205.2588}{{\tt arXiv:1205.2588}}.

\bibitem{Schicker:2014aoa}
R.~Schicker, {\it {Diffractive production of mesons}},
  \href{http://arxiv.org/abs/1410.6060}{{\tt arXiv:1410.6060}}.

\bibitem{Staszewski:2011bg}
R.~Staszewski, P.~Lebiedowicz, M.~Trzebi{\'n}ski, J.~Chwastowski, and
  A.~Szczurek, {\it {Exclusive $\pi^+ \pi^-$ Production at the LHC with Forward
  Proton Tagging}},  {\em Acta Phys.Polon.} {\bf B42} (2011) 1861--1870,
  [\href{http://arxiv.org/abs/1104.3568}{{\tt arXiv:1104.3568}}].

\bibitem{CMS_private_com}
D.~d'Enterria.
\newblock {Private communication}.

\bibitem{Osterberg:2014mta}
K.~{\"O}sterberg, {\it {Potential of central exclusive production studies in
  high $\beta^{*}$ runs at the LHC with CMS-TOTEM}},  {\em Int.J.Mod.Phys.}
  {\bf A29} (2014), no.~28 1446019.

\bibitem{Lebiedowicz:2009pj}
P.~Lebiedowicz and A.~Szczurek, {\it {Exclusive $pp \to pp \pi^{+}\pi^{-}$
  reaction: From the threshold to LHC}},  {\em Phys.Rev.} {\bf D81} (2010)
  036003, [\href{http://arxiv.org/abs/0912.0190}{{\tt arXiv:0912.0190}}].

\bibitem{Lebiedowicz:2011nb}
P.~Lebiedowicz, R.~Pasechnik, and A.~Szczurek, {\it {Measurement of exclusive
  production of scalar $\chi_{c0}$ meson in proton-(anti)proton collisions via
  $\chi_{c0} \to \pi^{+}\pi^{-}$ decay}},  {\em Phys.Lett.} {\bf B701} (2011)
  434--444, [\href{http://arxiv.org/abs/1103.5642}{{\tt arXiv:1103.5642}}].

\bibitem{Lebiedowicz:2011tp}
P.~Lebiedowicz and A.~Szczurek, {\it {$pp \to pp K^{+}K^{-}$ reaction at high
  energies}},  {\em Phys.Rev.} {\bf D85} (2012) 014026,
  [\href{http://arxiv.org/abs/1110.4787}{{\tt arXiv:1110.4787}}].

\bibitem{Lebiedowicz:thesis}
P.~Lebiedowicz, {\em {Exclusive reactions with light mesons: From low to high
  energies}}.
\newblock PhD thesis, {IFJ PAN}, {2014}.
\newblock The thesis is available at
  \url{http://www.ifj.edu.pl/msd/rozprawy_dr/rozpr_Lebiedowicz.pdf}.

\bibitem{Lebiedowicz:2013ika}
P.~Lebiedowicz, O.~Nachtmann, and A.~Szczurek, {\it {Exclusive central
  diffractive production of scalar and pseudoscalar mesons; tensorial vs.
  vectorial pomeron}},  {\em Annals Phys.} {\bf 344} (2014) 301--339,
  [\href{http://arxiv.org/abs/1309.3913}{{\tt arXiv:1309.3913}}].

\bibitem{Ewerz:2013kda}
C.~Ewerz, M.~Maniatis, and O.~Nachtmann, {\it {A Model for Soft High-Energy
  Scattering: Tensor Pomeron and Vector Odderon}},  {\em Annals of Physics}
  {\bf 342} (2014) 31--77, [\href{http://arxiv.org/abs/1309.3478}{{\tt
  arXiv:1309.3478}}].

\bibitem{Bolz:2014mya}
A.~Bolz, C.~Ewerz, M.~Maniatis, O.~Nachtmann, M.~Sauter, and A.~Sch{\"o}ning,
  {\it {Photoproduction of $\pi^{+} \pi^{-}$ pairs in a model with
  tensor-pomeron and vector-odderon exchange}},
  \href{http://arxiv.org/abs/1409.8483}{{\tt arXiv:1409.8483}}.

\bibitem{Sauter_LowX}
M.~Sauter.
\newblock {A talk \textit{Photoproduction of $\pi^{+}\pi^{-}$ pairs in a model
  with tensor-pomeron and vector-odderon exchange}, Low x Meeting, 17-21 June
  2014, Kyoto, Japan}.

\bibitem{Drell:1960zz}
S.~D. Drell, {\it {Production of Particle Beams at Very High Energies}},  {\em
  Phys.Rev.Lett.} {\bf 5} (1960) 278--281.

\bibitem{Drell:1961zz}
S.~D. Drell, {\it {Peripheral Contributions to High-Energy Interaction
  Processes}},  {\em Rev.Mod.Phys.} {\bf 33} (1961) 458--466.

\bibitem{Soding:1965nh}
P.~S{\"o}ding, {\it {On the apparent shift of the rho meson mass in
  photoproduction}},  {\em Phys.Lett.} {\bf 19} (1966) 702--704.

\bibitem{Szczurek:2004xe}
A.~Szczurek and A.~P. Szczepaniak, {\it {Diffractive photoproduction of
  opposite-charge pseudoscalar meson pairs at high energies}},  {\em Phys.Rev.}
  {\bf D71} (2005) 054005, [\href{http://arxiv.org/abs/hep-ph/0410083}{{\tt
  hep-ph/0410083}}].

\bibitem{Armesto:2014sma}
N.~Armesto and A.~H. Rezaeian, {\it {Exclusive vector meson production at high
  energies and gluon saturation}},  {\em Phys.Rev.} {\bf D90} (2014) 054003,
  [\href{http://arxiv.org/abs/1402.4831}{{\tt arXiv:1402.4831}}].

\bibitem{Santos:2014vwa}
G.~S. dos Santos and M.~V.~T. Machado, {\it {Light vector meson photoproduction
  in hadron-hadron and nucleus-nucleus collisions at the energies available at
  the CERN Large Hadron Collider}},  \href{http://arxiv.org/abs/1407.4148}{{\tt
  arXiv:1407.4148}}.

\bibitem{Schafer:2007mm}
W.~Sch{\"a}fer and A.~Szczurek, {\it {Exclusive photoproduction of $J/\psi$ in
  proton-proton and proton-antiproton scattering}},  {\em Phys.Rev.} {\bf D76}
  (2007) 094014, [\href{http://arxiv.org/abs/0705.2887}{{\tt
  arXiv:0705.2887}}].

\bibitem{Cisek:2010jk}
A.~Cisek, W.~Sch{\"a}fer, and A.~Szczurek, {\it {Exclusive photoproduction of
  $\phi$ meson in $\gamma p \to \phi p$ and $p p \to p \phi p$ reactions}},
  {\em Phys.Lett.} {\bf B690} (2010) 168--174,
  [\href{http://arxiv.org/abs/1004.0070}{{\tt arXiv:1004.0070}}].

\bibitem{Cisek:2011vt}
A.~Cisek, P.~Lebiedowicz, W.~Sch{\"a}fer, and A.~Szczurek, {\it {Exclusive
  production of $\omega$ meson in proton-proton collisions at high energies}},
  {\em Phys.Rev.} {\bf D83} (2011) 114004,
  [\href{http://arxiv.org/abs/1101.4874}{{\tt arXiv:1101.4874}}].

\bibitem{Cisek:2014ala}
A.~Cisek, W.~Sch{\"a}fer, and A.~Szczurek, {\it {Exclusive photoproduction of
  charmonia in $\gamma p \to V p$ and $p p \to p V p$ reactions within
  $k_t$-factorization approach}},  \href{http://arxiv.org/abs/1405.2253}{{\tt
  arXiv:1405.2253}}.

\bibitem{Lebiedowicz:2013vya}
P.~Lebiedowicz and A.~Szczurek, {\it {Exclusive $p p \to p p \pi^{0}$ reaction
  at high energies}},  {\em Phys.Rev.} {\bf D87} (2013) 074037,
  [\href{http://arxiv.org/abs/1303.2882}{{\tt arXiv:1303.2882}}].

\bibitem{Lebiedowicz:2013xlb}
P.~Lebiedowicz and A.~Szczurek, {\it {Exclusive diffractive photon
  bremsstrahlung at the LHC}},  {\em Phys.Rev.} {\bf D87} (2013) 114013,
  [\href{http://arxiv.org/abs/1302.4346}{{\tt arXiv:1302.4346}}].

\bibitem{Adler:2002sc}
{\bf STAR} Collaboration, C.~Adler et~al., {\it {Coherent $\rho^{0}$ Production
  in Ultraperipheral Heavy Ion Collisions}},  {\em Phys.Rev.Lett.} {\bf 89}
  (2002) 272302, [\href{http://arxiv.org/abs/nucl-ex/0206004}{{\tt
  nucl-ex/0206004}}].

\bibitem{Abelev:2007nb}
{\bf STAR} Collaboration, B.~I. Abelev et~al., {\it {$\rho^{0}$ Photoproduction
  in Ultraperipheral Relativistic Heavy Ion Collisions at $\sqrt{s_{NN}}$ = 200
  GeV}},  {\em Phys.Rev.} {\bf C77} (2008) 034910,
  [\href{http://arxiv.org/abs/0712.3320}{{\tt arXiv:0712.3320}}].

\bibitem{Abelev:2008ew}
{\bf STAR} Collaboration, B.~I. Abelev et~al., {\it {Observation of Two-Source
  Interference in the Photoproduction Reaction $Au Au \to Au Au \rho^{0}$}},
  {\em Phys.Rev.Lett.} {\bf 102} (2009) 112301,
  [\href{http://arxiv.org/abs/0812.1063}{{\tt arXiv:0812.1063}}].

\bibitem{Agakishiev:2011me}
{\bf STAR} Collaboration, G.~Agakishiev et~al., {\it {$\rho^{0}$
  Photoproduction in $Au Au$ Collisions at $\sqrt{s_{NN}}$ = 62.4 GeV with
  STAR}},  {\em Phys.Rev.} {\bf C85} (2012) 014910,
  [\href{http://arxiv.org/abs/1107.4630}{{\tt arXiv:1107.4630}}].

\bibitem{Nystrand:2014vra}
{\bf for the ALICE} Collaboration, J.~Nystrand, {\it {Photonuclear Production
  of Vector Mesons in Ultra-Peripheral $Pb$-$Pb$ Collisions at the LHC}},
  \href{http://arxiv.org/abs/1408.0811}{{\tt arXiv:1408.0811}}.

\bibitem{Baur:2001jj}
G.~Baur, K.~Hencken, D.~Trautmann, S.~Sadovsky, and Y.~Kharlov, {\it {Coherent
  $\gamma \gamma$ and $\gamma A$ Interactions in Very Peripheral Collisions at
  Relativistic Ion Colliders}},  {\em Phys.Rept.} {\bf 364} (2002) 359--450,
  [\href{http://arxiv.org/abs/hep-ph/0112211}{{\tt hep-ph/0112211}}].

\bibitem{Bertulani:2005ru}
C.~A. Bertulani, S.~R. Klein, and J.~Nystrand, {\it {Physics of
  Ultra-Peripheral Nuclear Collisions}},  {\em Ann.Rev.Nucl.Part.Sci.} {\bf 55}
  (2005) 271--310, [\href{http://arxiv.org/abs/nucl-ex/0502005}{{\tt
  nucl-ex/0502005}}].

\bibitem{Baltz:2007kq}
A.~J. Baltz, G.~Baur, D.~d'Enterria, L.~Frankfurt, F.~Gelis, et~al., {\it {The
  Physics of Ultraperipheral Collisions at the LHC}},  {\em Phys.Rept.} {\bf
  458} (2008) 1--171, [\href{http://arxiv.org/abs/0706.3356}{{\tt
  arXiv:0706.3356}}].

\bibitem{Donnachie:2002en}
A.~Donnachie, H.~G. Dosch, P.~V. Landshoff, and O.~Nachtmann, {\it {Pomeron
  physics and QCD}},  {\em Camb.Monogr.Part.Phys.Nucl.Phys.Cosmol.} {\bf 19}
  (2002) 1--347.

\bibitem{Agashe:2014kda}
{\bf Particle Data Group} Collaboration, K.~A. Olive et~al., {\it {Review of
  Particle Physics}},  {\em Chin.Phys.} {\bf C38} (2014) 090001.

\bibitem{Friman:1995qm}
B.~Friman and M.~Soyeur, {\it {Photoproduction of vector mesons off nucleons
  near threshold}},  {\em Nucl.Phys.} {\bf A600} (1996) 477--490,
  [\href{http://arxiv.org/abs/nucl-th/9601028}{{\tt nucl-th/9601028}}].

\bibitem{Laget:2000gj}
J.~M. Laget, {\it {Photoproduction of vector mesons at large momentum
  transfer}},  {\em Phys.Lett.} {\bf B489} (2000) 313--318,
  [\href{http://arxiv.org/abs/hep-ph/0003213}{{\tt hep-ph/0003213}}].

\bibitem{Oh:2003gm}
Y.~Oh, {\it {Vector meson photoproduction processes near threshold}},  {\em
  J.Korean Phys.Soc.} {\bf 43} (2003) S20--S26,
  [\href{http://arxiv.org/abs/nucl-th/0301011}{{\tt nucl-th/0301011}}].

\bibitem{Oh:2003aw}
Y.~Oh and T.-S.~H. Lee, {\it {$\rho$ meson photoproduction at low-energies}},
  {\em Phys.Rev.} {\bf C69} (2004) 025201,
  [\href{http://arxiv.org/abs/nucl-th/0306033}{{\tt nucl-th/0306033}}].

\bibitem{Riek:2008ct}
F.~Riek, R.~Rapp, T.-S.~H. Lee, and Y.~Oh, {\it {Medium Effects in $\rho$-Meson
  Photoproduction}},  {\em Phys.Lett.} {\bf B677} (2009) 116--120,
  [\href{http://arxiv.org/abs/0812.0987}{{\tt arXiv:0812.0987}}].

\bibitem{Obukhovsky:2009th}
I.~T. Obukhovsky, A.~Faessler, D.~K. Fedorov, T.~Gutsche, V.~E. Lyubovitskij,
  V.~G. Neudatchin, and L.~L. Sviridova, {\it {Quasielastic $\rho^0$
  electroproduction on the proton at intermediate energy: Role of scalar and
  pseudoscalar meson exchange}},  {\em Phys.Rev.} {\bf D81} (2010) 013007,
  [\href{http://arxiv.org/abs/0911.3074}{{\tt arXiv:0911.3074}}].

\bibitem{Breitweg:1997ed}
{\bf ZEUS} Collaboration, J.~Breitweg et~al., {\it {Elastic and proton
  dissociative $\rho^0$ photoproduction at HERA}},  {\em Eur.Phys.J.} {\bf C2}
  (1998) 247--267, [\href{http://arxiv.org/abs/hep-ex/9712020}{{\tt
  hep-ex/9712020}}].

\bibitem{Breitweg:1999jy}
{\bf ZEUS} Collaboration, J.~Breitweg et~al., {\it {Measurement of diffractive
  photoproduction of vector mesons at large momentum transfer at HERA}},  {\em
  Eur.Phys.J.} {\bf C14} (2000) 213--238,
  [\href{http://arxiv.org/abs/hep-ex/9910038}{{\tt hep-ex/9910038}}].

\bibitem{Derrick:1994dt}
{\bf ZEUS} Collaboration, M.~Derrick et~al., {\it {Measurement of total and
  partial photon proton cross-sections at 180 GeV center of mass energy}},
  {\em Z.Phys.} {\bf C63} (1994) 391--408.

\bibitem{Derrick:1995vq}
{\bf ZEUS} Collaboration, M.~Derrick et~al., {\it {Measurement of elastic
  $\rho^0$ photoproduction at HERA}},  {\em Z.Phys.} {\bf C69} (1995) 39--54,
  [\href{http://arxiv.org/abs/hep-ex/9507011}{{\tt hep-ex/9507011}}].

\bibitem{Derrick:1996yt}
{\bf ZEUS} Collaboration, M.~Derrick et~al., {\it {Measurement of elastic
  $\omega$ photoproduction at HERA}},  {\em Z.Phys.} {\bf C73} (1996) 73--84,
  [\href{http://arxiv.org/abs/hep-ex/9608010}{{\tt hep-ex/9608010}}].

\bibitem{Aid:1996bs}
{\bf H1} Collaboration, S.~Aid et~al., {\it {Elastic photoproduction of
  $\rho^0$ mesons at HERA}},  {\em Nucl.Phys.} {\bf B463} (1996) 3--32,
  [\href{http://arxiv.org/abs/hep-ex/9601004}{{\tt hep-ex/9601004}}].

\bibitem{Melikhov:2003hs}
D.~Melikhov, O.~Nachtmann, V.~Nikonov, and T.~Paulus, {\it {Masses and
  couplings of vector mesons from the pion electromagnetic, weak, and $\pi
  \gamma$ transition form factors}},  {\em Eur.Phys.J.} {\bf C34} (2004)
  345--360, [\href{http://arxiv.org/abs/hep-ph/0311213}{{\tt hep-ph/0311213}}].

\bibitem{Poppe:1986dq}
M.~Poppe, {\it {Exclusive Hadron Production in Two Photon Reactions}},  {\em
  Int.J.Mod.Phys.} {\bf A1} (1986) 545--668.

\bibitem{Szczurek:2002bn}
A.~Szczurek and J.~Speth, {\it {Perturbative QCD versus pion exchange and
  hadronic FSI effects in the $\gamma \gamma \to \pi^{+} \pi^{-}$ reaction}},
  {\em Nucl.Phys.} {\bf A728} (2003) 182--202,
  [\href{http://arxiv.org/abs/hep-ph/0207265}{{\tt hep-ph/0207265}}].

\bibitem{Klusek-Gawenda:2013rtu}
M.~K{\l}usek-Gawenda and A.~Szczurek, {\it {$\pi^+ \pi^-$ and $\pi^0 \pi^0$
  pair production in photon-photon and in ultraperipheral ultrarelativistic
  heavy ion collisions}},  {\em Phys.Rev.} {\bf C87} (2013), no.~5 054908,
  [\href{http://arxiv.org/abs/1302.4204}{{\tt arXiv:1302.4204}}].

\end{thebibliography}\endgroup
\end{small}
}
%------------------------------------------------------------------

\end{document}